\documentclass[amsmath,amssym,ams, preprint, superscriptaddress,nofootinbib]{article}
\usepackage[a4paper,textwidth=17.9cm,textheight=26cm]{geometry}
\usepackage[english]{babel}
\usepackage[explicit]{titlesec}
\renewcommand{\baselinestretch}{1.5} 
\usepackage[T1]{fontenc}
\usepackage{lmodern}
\usepackage{helvet}
\usepackage{lineno}
\usepackage{graphicx}
\usepackage{xr}
\usepackage{authblk}
\usepackage{ulem}
\usepackage{siunitx}
\usepackage{upgreek}
\usepackage{amsmath}
\usepackage{booktabs}
\usepackage{tabularx}
\usepackage{makecell}
\usepackage{threeparttable}
\usepackage{array}

\newcolumntype{Y}{>{\raggedright\arraybackslash}X}

\newcommand{\nocontentsline}[3]{}
\newcommand{\tocless}[2]{\bgroup\let\addcontentsline=\nocontentsline#1{#2}\egroup}

\makeatletter
\newcommand*{\addFileDependency}[1]{
  \typeout{(#1)}
  \@addtofilelist{#1}
  \IfFileExists{#1}{}{\typeout{No file #1.}}
}
\makeatother

\addto\captionsenglish{}
\addto\captionsenglish{}

\begin{document}

\title{Hz-resolution wide-span photonic integrated terahertz signal analyzer}
\author[1, 2]{Aleksei Gaier}
\author[1, 2]{Jiawen Liu}
\author[1]{Ekaterina Mironova}
\author[1, 2]{Gabriel Jülg}
\author[1, 2]{Ileana-Cristina Benea-Chelmus}

\affil[1]{Hybrid Photonics Laboratory, École Polytechnique Fédérale de Lausanne (EPFL),  CH-1015, Switzerland}
\affil[2]{Center for Quantum Science and Engineering (QSE), CH-1015, Switzerland}

\date{\today}
\maketitle

\begin{abstract}
Wide-span spectral and noise characterization at millimeter-wave and terahertz frequencies is increasingly important for emerging wireless, sensing, and spectroscopy systems, yet remains challenging for conventional electronic instrumentation because of the complexity and calibration burden of extender-, multiplier-, and mixer-chain architectures. Here we show that photonics can provide attractive alternatives to this highly challenging electronic instrumentation through an antenna-coupled thin-film lithium niobate electro-optic receiver. Our implementation performs fast spectral reconstruction across the widely separated WR 9.0 (80-125 GHz) and WR 2.8 (240-380 GHz) bands with a single component, with carrier-frequency errors below 3~MHz, scan speeds up to 25~THz/s, and a displayed average noise level below $-104$~dBm/Hz. We then extend to Hz-level resolution and demonstrate phase-noise characterisation capabilities across these ultra-wide bands by using a mode-locked femtosecond-laser frequency comb as a sampling clock, mapping high-frequency carriers to aliased intermediate-frequency tones. We validate the method by measuring the phase noise of 90.225~GHz and 270.675~GHz carriers using the same chip and system configuration. We further improve the sensitivity of the phase-noise measurement through cross-correlation between two optical probe-pulse channels. Finally, we show that the same technique can be applied synchronously to ten widely spaced frequencies distributed across more than 100~GH, with the option to quantify their mutual coherence. These results establish TFLN integrated photonics as a scalable route toward compact millimeter-wave/terahertz instrumentation.
\end{abstract}

\maketitle

\tocless\section{Introduction}\label{sec_intro}

\begin{figure}[h!]
    \centering
    \includegraphics{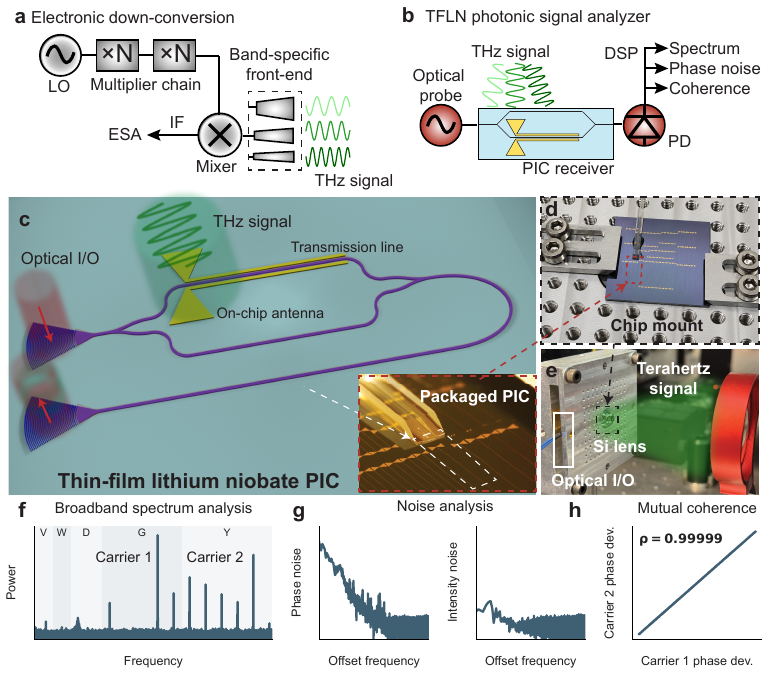}
    \renewcommand{\baselinestretch}{1} 
    \caption{\textbf{System-level concept and functionalities of the TFLN-based mmWave/THz signal analyzer}
\textbf{(a)} Conventional electronic mmWave/THz signal analysis relies on band-specific front-ends, LO multiplication, and mixer-based down-conversion to an intermediate-frequency (IF) signal. A broadband signal containing frequency components distributed over multiple waveguide bands would therefore require several front-end configurations or sequential measurements with different mixer heads before analysis by an electronic signal analyzer (ESA).
\textbf{(b)} In this work, incident mmWave/THz fields are transduced to the optical domain using a single antenna-coupled thin-film lithium niobate (TFLN) photonic integrated circuit (PIC). The same photonic front-end can receive multi-component signals and, after photodetection (PD) and digital signal processing (DSP), provide access to spectral amplitudes, phase noise, and mutual coherence.
\textbf{(c)} Operating principle of the TFLN PIC. An incident mmWave/THz signal is received by the on-chip antenna and guided by the coplanar-strip transmission line placed alongside the optical waveguide, where the electric field modulates the optical carrier via the Pockels effect. The modulated optical signal is read out through fiber-coupled optical input/output.
\textbf{(d,e)} Photographs of the chip mount and free-space THz interface, including the optical I/O and silicon lens.
\textbf{(f)} Broadband spectral analysis spanning multiple standard waveguide bands.
\textbf{(g)} Phase- and intensity-noise observables recovered from the detected time-domain signal.
\textbf{(h)} Mutual-coherence analysis between two frequency components, represented by the correlation of their recovered phase deviations.}

    \renewcommand{\baselinestretch}{1.5} 
    \label{fig1}  
\end{figure}

Over the past decades, electronic instrumentation and integrated circuits have achieved remarkable progress toward millimeter (mmWave) and sub-terahertz (THz) operation, enabling increasingly capable measurement systems \cite{choi2024acsphotonics, memioglu2021,hara2018,hu2018rfic,ziegler300fmcw}. Nevertheless, spectrum and signal analysis become progressively more challenging as carrier frequencies approach the sub-THz regime. At these wavelengths, receiver front-ends, frequency extenders, and mixers are commonly implemented with metallic waveguide interfaces whose dimensions define finite operating bands. As a result, extending frequency coverage typically requires band-specific waveguide front-ends, multiplier-chain local oscillators, and mixer-based down-conversion modules (Fig.~\ref{fig1}a). A signal containing components distributed across multiple waveguide bands, therefore, requires several front-end configurations, separate calibrations and sequential measurements. This hardware segmentation is further compounded by frequency-dependent performance degradation: at higher carrier frequencies, conversion loss, system complexity, and calibration overhead generally increase \cite{sengupta2018natelectronics}. In addition, phase noise and spurious content generally scale unfavorably with multiplication factor, degrading effective sensitivity and measurement uncertainty at higher carrier frequencies. In essence, wide-coverage instrumentation remains constrained by segmented hardware, limited instantaneous bandwidth, and the need for low-noise references throughout the conversion chain.

Beyond simply resolving spectral amplitudes, many emerging mmWave/THz sources and systems rely critically on mutual coherence between frequency components separated by hundreds of gigahertz or more, often spanning different waveguide bands. This makes the measurement problem even more demanding: the instrument must not only detect widely separated components synchronously, but also preserve their relative phase across the measurement bandwidth. This measurement problem is particularly evident in the rapid development of terahertz frequency combs, including electrically pumped THz quantum-cascade-laser (QCL) combs \cite{burghoff2014thzcomb,senica2022planarized}. In comb-based THz spectroscopy, multi-heterodyne and dual-comb techniques map optical- or THz-frequency information onto radio-frequency beat notes; this mapping is only faithful if the phases of the comb lines remain correlated across the spectrum over the measurement time \cite{picque2019natphotonics,yang2016optica_thzmultiheterodyne,consolino2020hybriddualcomb}. Decoherence directly broadens multi-heterodyne beats and reduces spectral resolution\cite{djevahirdjian2023_natcom_thz,shin2023_natcom_combrooted,consolino2019phase_stabilized_qcl}. Similarly, in wideband mmWave/THz communications, maintaining stable relative phase between carriers is essential for coherent multi-carrier transmission, and decoherence across the band can directly limit link performance \cite{jia20202, shams2014photonic,nagatsuma2016natphotonics}.

These examples expose a key limitation of conventional electronic approaches: they must not only detect widely separated sub-THz and THz spectrum components, but also preserve their relative phase through frequency conversion, band-limited front-ends, and separate calibration paths. Photonics offers an attractive alternative by transferring THz information to the optical domain, where stable laser references, frequency combs, large optical bandwidths, and fiber-compatible routing can be exploited \cite{picque2019natphotonics,fortier2019commsphys,parriaux2020aop}. Early and recent demonstrations have realized THz spectrum analysis using optoelectronic receivers and comb-referenced and optoelectronic heterodyne concepts \cite{yasui2009oe_thzcombsa,yee2010ol_thzsa,fernandezolvera2021_ieeeaccess_trueosa,krause2022_ieeeaccess_psa,theis2023_tthz_hybrid, preu2025continuous, krause2025natcom}. In particular, comb-assisted techniques have enabled precision metrology of THz QCLs, including frequency-noise and intrinsic-linewidth measurements, unambiguous real-time absolute-frequency readout, and direct electric-field sampling of free-running THz QCL combs \cite{ravaro2012oe,kliebisch2018unambiguous,markmann2023electro}. Related work on frequency-stable THz synthesis and distribution further highlights the strengths of photonic referencing for precision spectroscopy and metrology \cite{mouret2009_oe_photomixing}. These demonstrations show that photonic referencing can provide precise access to THz frequency, phase, and field dynamics. However, most existing implementations rely on bulk free-space laboratory systems and are optimized for specific source-metrology tasks. Moving from such laboratory-scale photonic metrology toward compact and deployable instruments, therefore, calls for integrated platforms that combine chip-scale detector heads, stable optical referencing, fiber-compatible readout, and wide-span coupling to mmWave/THz signals. 

Thin-film lithium niobate (TFLN) integrated photonics is particularly well suited to this role, owing to its strong and intrinsically broadband Pockels effect, low optical loss, and strong field confinement in optical waveguides \cite{boes2018science_ln,mercante2018oe,zhang2022prj}. Building on these properties, recent demonstrations have positioned TFLN as a strong platform for THz systems, including integrated multi-THz-bandwidth detection \cite{herter2025natcom_thzdet}, beam-profiling \cite{tomasino2026lsa} and generation \cite{lampert2025natcom,zhang2025natcom_ln_thz} architectures. In parallel, TFLN has enabled integrated mmWave-photonic functionalities, including photonic receivers and radar architectures \cite{mollerdefreitas2025,zhu2025natphotonics_radar,tao2025ultrabroadband,zhang2026integrated}.

Here, we extend the role of TFLN photonics toward chip-scale mmWave/THz signal analysis by demonstrating a spectrum-analysis and phase-noise/coherence-measurement system based on an antenna-coupled TFLN electro-optic receiver (Fig.~\ref{fig1}b-e). Incident mmWave/THz fields are encoded onto an optical carrier and retrieved by photodetection and digital signal processing. The use of a rapidly swept external-cavity laser enables high-speed spectral acquisition across multiple waveguide bands with scan speeds up to 25 THz/s, while additionally implementing a frequency comb provides a stable optical-frequency ruler for accurate extraction of the incident THz frequencies. Beyond spectral amplitude, we extract phase fluctuations to measure single-sideband phase noise and quantify mutual coherence between widely separated frequency components, enabling correlation-based analysis over broad frequency spans. These results position TFLN integrated photonics as a scalable path toward compact mmWave/THz spectrum and noise analyzers that mitigate key limitations of purely electronic multiplier-chain instrumentation.

\tocless\section{Results}\label{sec_results}
\tocless\subsection{Fast broadband spectrum analysis with an antenna-coupled TFLN PIC}
\begin{figure}[h!]
    \centering
    \includegraphics[width=18cm]{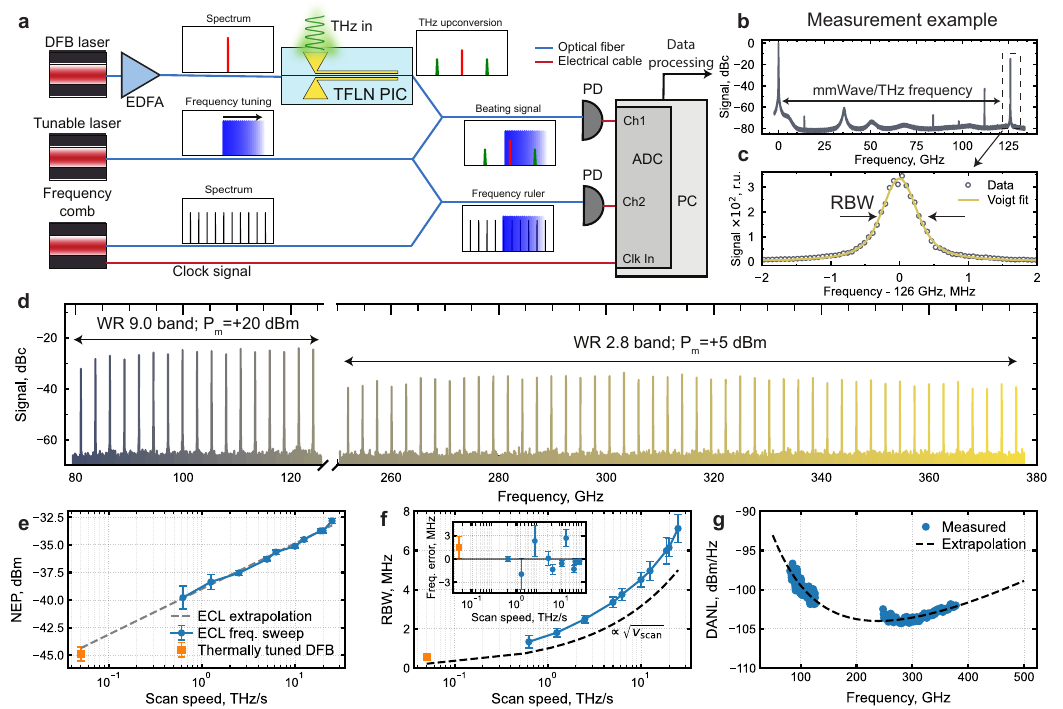}
    \renewcommand{\baselinestretch}{1} 
    \caption{\textbf{Fast broadband mmWave/THz spectrum analysis with a comb-referenced antenna-coupled TFLN photonic receiver.}
    \textbf{(a)} Measurement setup. A CW mmWave/THz signal is received by a broadband bowtie antenna and coupled to an on-chip transmission line surrounding the TFLN optical waveguide. The optical pump is amplified by an erbium-doped fiber amplifier (EDFA) before the chip to increase the sideband power generated through electro-optic interaction. A tunable laser is swept across the sideband spectrum, and the resulting heterodyne beat note is detected on a photodiode (PD). In parallel, the swept laser is heterodyned with an optical frequency comb, providing an optical frequency ruler that calibrates the instantaneous laser detuning during the scan.
    \textbf{(b)} Representative reconstructed broadband spectrum.
    \textbf{(c)} Zoom-in of an individual spectral line near 126~GHz. The line is fitted with a Voigt profile to extract the center frequency and resolution bandwidth (RBW).
    \textbf{(d)} Reconstructed single-tone spectra across the WR,9.0 and WR,2.8 waveguide bands, measured by stepping the mmWave/THz source frequency while using the same photonic readout architecture. The spectra were acquired at a scan speed of 25~THz/s.
    \textbf{(e)} Noise-equivalent power (NEP) at 126~GHz as a function of optical scan speed. Consistent NEP values are obtained using both swept external-cavity-laser operation and thermally tuned DFB operation, demonstrating that the photonic readout remains stable and reproducible across different laser-tuning architectures. 
    \textbf{(f)} Extracted RBW as a function of optical scan speed. The dashed curve indicates the expected $\sqrt{v_\mathrm{scan}}$ scaling, where $v_\mathrm{scan}$ is the scan speed. The close agreement confirms scan speed-limited spectral resolution. The inset shows the corresponding frequency readout error, defined as the difference between the set frequency (126 GHz) and the measured frequency. 
    \textbf{(g)} Displayed average noise level (DANL), defined as NEP normalized to a 1~Hz bandwidth, across the measured frequency range together with an extrapolated response.
    }
    \renewcommand{\baselinestretch}{1.5} 
    \label{fig2}  
\end{figure}

We first demonstrate fast, wide-span mmWave/THz spectrum analysis using the antenna-coupled TFLN PIC as a compact electro-optic front end with a geometry similar to our recent work~\cite{gaier2025wireless}. The device fabrication process is described in Methods~\ref{fabrication}4.1. Fig.~\ref{fig2} presents the system concept and experimental setup (additional details on the setup are provided in the Methods section~\ref{setupdetails_fast}4.2). The PIC employs a broadband bowtie antenna that couples incident continuous-wave (CW) mmWave/THz radiation into an on-chip transmission line surrounding the optical waveguide. The transmission-line geometry was designed to balance broadband antenna--transmission-line coupling, optical propagation loss, and electro-optic field overlap in the interaction region.

A distributed feedback pump laser (DFB) is amplified by an erbium-doped fiber amplifier (EDFA) and coupled to the TFLN PIC exposed to incident mmWave/THz radiation (Fig.~\ref{fig2}a). The generated electro-optic sidebands are read out by sweeping a second tunable laser across the sideband spectrum and detecting the heterodyne beat note on a photodiode. Unlike our previous work which lacked a precise frequency ruler~\cite{gaier2025wireless}, in this case, a fraction of the swept laser is simultaneously heterodyned with an optical frequency comb to calibrate the laser detuning during the scan. The optical frequency comb, therefore, acts as an absolute frequency ruler for the swept readout, enabling accurate mmWave/THz frequency extraction even during rapid and nonlinear laser scans. Additional details are provided in the supplementary information~\ref{SInote2}. The system was operated free-running, without active laser-to-comb locking, to prioritize maximum scan speed.

Fig.~\ref{fig2}b,c shows a representative broadband scan and a zoom-in of an individual spectral line, fitted with a Voigt profile to extract its center frequency and resolution bandwidth. Repeating this measurement while stepping the mmWave/THz source frequency, we reconstruct single-tone spectra across the WR\,9.0 and WR\,2.8 bands (Fig.~\ref{fig2}d). Together with the broadband antenna--transmission-line response predicted in supplementary information~\ref{PIC_design}, these measurements indicate a route toward operation at still higher carrier frequencies.

Owing to the implemented comb-based frequency ruler, the measurement remains robust over optical scan speeds up to 25~THz/s where the otherwise strong deviations of the frequency tuning from a linear frequency sweep would significantly worsen the frequency precision (Fig.~\ref{fig2}f). This rapid acquisition of broadband mmWave/THz spectra is well above the sweep rates typically available in electronically swept mmWave/THz spectrum-analysis workflows (see Table 1), highlighting the advantage of optical-domain frequency acquisition. The scan speed directly affects the measured linewidth and sensitivity. At higher scan speeds, the swept laser spends less time within the detection bandwidth of a given electro-optic sideband. As a result, the reconstructed spectral line broadens, increasing the effective resolution bandwidth (RBW). Since the detected noise is integrated over a larger bandwidth, the noise power in each frequency bin increases, thereby increasing the noise-equivalent power (NEP), defined as the mmWave/THz power level at which the signal-to-noise ratio becomes 1 for a given scan speed. Overall, there is trade-off between faster scans, providing higher acquisition speed, and higher NEP and larger RBW. The extracted RBW follows the expected scan-speed-limited broadening, indicating that the spectral resolution is primarily set by the optical sweep speed rather than by additional instrumental broadening (see supplementary information~\ref{si-resolution}). At the same time, the absolute frequency error remains low and comparable to the extracted RBW over the investigated scan-speed range (inset Fig.~\ref{fig2}f). Importantly, the extracted NEP remains at least one order of magnitude below the power levels typically available from coherent CW mmWave/THz sources, including Schottky-diode frequency-multiplier chains~\cite{zhang2022frequency_multipliers}, voltage-controlled oscillators~\cite{kraus2024dband_vco}, and photonic transmitters based on uni-traveling-carrier photodiodes (UTC-PDs)~\cite{heffernan2023microcomb_utcpd}, suggesting the compatibility of our instrument with these standard sources.

Fig.~\ref{fig2}g summarizes the displayed average noise level (DANL) over the measured frequency range. Since the measured NEP depends on the resolution bandwidth, which itself changes with optical scan speed, the DANL is reported as the RBW-normalized NEP, $\mathrm{DANL}=\mathrm{NEP}/\mathrm{RBW}$. This metric provides a scan-condition-independent measure of the receiver sensitivity and therefore summarizes the intrinsic performance of the photonic spectrum-analysis approach. The DANL remains comparatively flat from the WR 9.0 to the WR 2.8 band, consistent with the broadband antenna-coupled transmission-line design and supporting the use of a single photonic readout architecture over widely separated mmWave/THz bands. The present sensitivity is limited in part by the fiber-to-fiber optical coupling efficiency of the chip, approximately $-23$~dB, and by residual optical noise reaching the photodiode. Supplementary Note~\ref{si2_snr} discusses the various noise contributions and strategies to mitigate them.

\tocless\subsection{Hz-resolution and phase-noise analysis through on-chip electro-optic sampling}
\begin{figure}[h!]
    \centering
    \includegraphics[width=18cm]{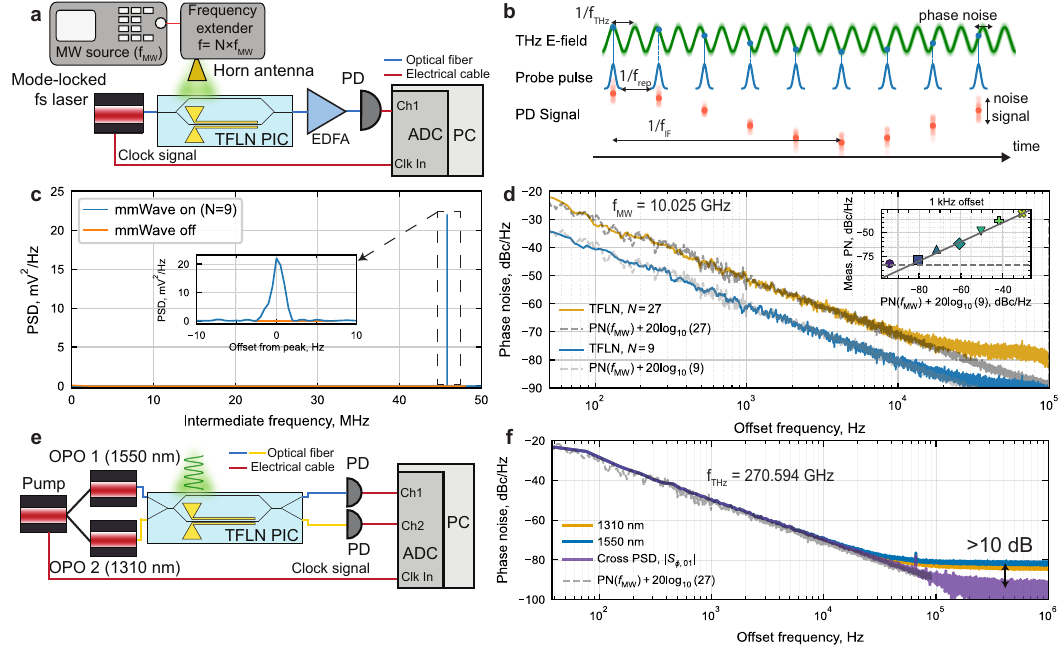}
    \renewcommand{\baselinestretch}{1} 
    \caption{\textbf{Hz-resolution and phase-noise analysis using on-chip electro-optic sampling.}
    \textbf{(a)} Measurement setup. A mode-locked fs laser probes the antenna-coupled TFLN PIC at repetition rate $f_{\mathrm{rep}}$. The incident mmWave/THz electric field modulates the optical pulse train through the Pockels effect; the MZI converts the induced phase modulation into intensity modulation, which is detected on a photodiode. An EDFA is used at the output to enhance the signal-to-noise ratio.
    \textbf{(b)} Optical sampling principle. Because the fs pulses are much shorter than the mmWave/THz period, each pulse samples the instantaneous electric field, preserving the carrier phase information in the pulse-to-pulse modulation. The high-frequency carrier is mapped to an aliased intermediate frequency $f_{\mathrm{IF}}=|f-kf_{\mathrm{rep}}|$, from which the absolute carrier frequency is recovered.
    \textbf{(c)} Measured electrical spectrum for $f_{\mathrm{MW}}=10.025$~GHz and multiplication factor $N=9$, showing the aliased carrier tone with the mmWave source on and off. The inset shows a zoom-in around the carrier peak, demonstrating sub-Hz level resolution.
    \textbf{(d)} Extracted single-sideband phase-noise spectra for carriers generated with multiplication factors $N=9$ and $N=27$ ($f_{\mathrm{MW}}=10.025$~GHz), together with the corresponding RF-source reference phase noise shifted by $20\log_{10}(N)$. Inset: measured phase-noise level at 1~kHz offset versus the expected multiplied reference level for different programmed phase-noise settings.
    \textbf{(e)} Two-channel cross-correlation measurement setup. Two synchronized optical pulse trains at 1550 and 1310 nm are generated by optical parametric oscillators (OPOs) pumped by the same fs laser and coupled into separate inputs of the TFLN PIC via wavelength-selective grating couplers. The two channels are combined on chip using Y-couplers and probe the same antenna-coupled electro-optic receiver. The spectral selectivity of the grating couplers provides additional isolation between the 1310 and 1550~nm channels. After the chip, both colors are detected by photodiodes and digitized simultaneously.
    \textbf{(f)} Single-channel and cross-correlated phase-noise spectra measured for a carrier at $f_{\mathrm{THz}}=270.594$~GHz. The individual 1310~nm and 1550~nm channels reach distinct readout floors at large offset frequencies, whereas the magnitude of the cross phase-noise PSD, $|S_{\phi,01}|$, rejects their uncorrelated noise contributions by more than 12~dB. The dashed trace shows the RF-source reference phase noise shifted by $20\log_{10}(27)$.}
    \renewcommand{\baselinestretch}{1.5} 
    \label{fig3}  
\end{figure}

Having established fast broadband mmWave/THz spectral readout, we next extend to Hz-resolution and demonstrate phase-noise analysis of mmWave/THz carriers using the antenna-coupled TFLN PIC receiver operated in a Mach-Zehnder interferometer (MZI) configuration (Fig.~\ref{fig3}a). As sketched in Fig.~\ref{fig3}b, a mode-locked fs laser probes the mmWave/terahertz signals captured by the TFLN PIC with a repetition rate $f_{\mathrm{rep}}\approx100$~MHz (additional details on the setup are provided in the Methods section~\ref{setupdetails_PN}4.3). Since the optical pulse duration is much shorter than the period of the mmWave/THz carrier, each pulse samples the local electric field at a well-defined time. The Pockels-induced phase shift of each pulse, therefore, depends on the carrier phase at the sampling instant, and the MZI converts this pulse-to-pulse phase modulation into intensity modulation (Fig.~\ref{fig3}b). Photodetection of the resulting pulse train produces an electrical signal in which a mmWave/THz carrier at frequency $f$ is mapped to an aliased intermediate frequency
\begin{equation}
f_{\mathrm{IF}} = \left| f - k f_{\mathrm{rep}} \right|,
\end{equation}
where the integer $k$ is such that $f_{\mathrm{IF}}\in[0,f_{\mathrm{rep}}/2]$. 

To enhance the detected signal-to-noise ratio, the optical signal was amplified after the chip using an EDFA (see Supplementary Information~\ref{si3_snr_pulsed}). We also note that although the broadband measurements in Fig.~\ref{fig2} employed direct phase-modulation readout to maximize sensitivity, the MZI receiver used here can also perform fast broadband spectral acquisition, albeit with an approximately 6~dB lower modulation response, as discussed in Supplementary Information~\ref{PIC_design}. 

The alias order $k$ can be retrieved from the coarse carrier-frequency estimate shown in Fig.~\ref{fig2} which, importantly, provides a resolution much below the repetition rate of the sampling laser (3 MHz resolution versus 100~MHz repetition rate), allowing unambiguous identification of $k$. Once $k$ is retrieved, the absolute mmWave/THz frequency $\hat{f}$ can be reconstructed from the measured $f_{\mathrm{IF}}$ and the comb repetition rate as
\begin{equation}
\hat{f} = k f_{\mathrm{rep}} \pm f_{\mathrm{IF}},
\end{equation}
with the sign determined from the coarse frequency estimate. Thus, the comb-referenced acquisition provides a precise carrier-frequency readout in the aliased domain while avoiding the need for an electronic local oscillator at the mmWave/THz carrier frequency. For the measurement shown in Fig.~\ref{fig3}c, the microwave seed frequency was $f_{\mathrm{MW}}=10.025$~GHz. After multiplication by $N=9$, this corresponds to a carrier at 90.225~GHz, measured at an aliased intermediate frequency of $f_{\mathrm{IF}}\approx45,817,289.7 \pm 0.5$~Hz for $f_{\mathrm{rep}}=99,967,682.4 \pm 0.2$~Hz and $k=903$. The reconstructed absolute carrier frequency is $\hat{f} = 90,224,999,917 \pm 181$~Hz. 

The detected spectrum in Fig.~\ref{fig3}c shows a Hz-linewidth (see inset) aliased carrier when the mmWave source is on, whereas the source-off trace remains at the noise floor. We then perform digital I/Q demodulation of this aliased frequency to extract the phase fluctuations and calculate the corresponding single-sideband phase-noise spectrum. Additional details on the data treatment are provided in Supplementary Information~\ref{SInote3}.2 ~\ref{SInote3}.3. The retrieval of amplitude noise, expressed as relative intensity noise (RIN), is discussed in Supplementary Information~\ref{AM_modulation}.

To quantitatively validate the phase-noise extraction, we generated carriers with calibrated phase-noise level of -79 dBc/Hz at 1 kHz offset using a Vector Signal Generator as the RF input before frequency multiplication. The RF frequency was set to $f_{\mathrm{MW}}=10.025$~GHz and multiplied by factors $N=9$ and $N=27$, yielding carriers at 90.225~GHz and 270.675~GHz, respectively. For $f_{\mathrm{rep}}=99.967$~MHz, these carriers correspond to aliased intermediate frequencies of approximately 45.817~MHz and 37.624~MHz, respectively. In both cases, the same MZI-based TFLN receiver and acquisition chain were used, without changing the receiver hardware or generating a tunable electronic LO at the carrier frequency.

Fig.~\ref{fig3}d compares the extracted phase-noise spectra for the $N=9$ and $N=27$ carriers. The measured spectra follow the expected reference curves obtained by adding $20\log_{10}(N)$ to the phase noise of the 10.025~GHz RF seed, confirming that the imposed phase fluctuations are transferred through the multiplier chain and correctly recovered by the TFLN receiver. As a final validation, the phase noise of the 10.025~GHz seed was measured both with an electrical spectrum analyzer and with our TFLN chip, showing excellent agreement in absolute units and confirming the accuracy of our calibration (see supplementary information~\ref{ESA_validation}). Finally, we sweep the programmed RF-source noise settings between -110 and -50 dBc/Hz at 1 kHz offset to test the dynamic range of the phase noise measurement. The inset of Fig.~\ref{fig3}d summarizes the extracted phase-noise level at a 1~kHz offset. The measured values agree with the expected multiplied reference levels over the investigated range, demonstrating quantitative phase-noise retrieval after large frequency multiplication. 

To further suppress the measurement noise floor, we get inspiration from the cross-correlation approach widely used in electronic phase-noise metrology~\cite{walls1992_crosscorrelation,rubiola2000_correlation} and translate it to photonics. As shown in Fig.~\ref{fig3}e, two synchronized pulse trains at 1550 and 1310~nm are generated by optical parametric oscillators pumped by the same femtosecond laser. The two wavelength channels are launched through dedicated grating couplers into separate input waveguides of the TFLN PIC and combined on chip using Y-couplers, such that both probe pulses experience the same electro-optic modulation induced by the incident mmWave/THz field. After electro-optic sampling, the two wavelength channels are routed to independent output paths. The two outputs are subsequently detected by two photodiodes and acquired simultaneously on two ADC channels. After digital I/Q demodulation, the retrieved phase fluctuations can be written as $\phi_{0,1}(t)=\phi_{\mathrm{IF}}(t)+n_{0,1}(t)$, where $\phi_{\mathrm{IF}}(t)$ contains the phase fluctuations common to both optical probes, whereas $n_{0,1}(t)$ represent the independent noise contributions of the respective photodetection and acquisition paths. The cross-spectrum is evaluated for each of $M$ successive time segments of a long acquisition, and then averaged as
\begin{equation}
    \hat{S}_{\phi,01}(f)
    =
    \frac{1}{M}
    \sum_{m=1}^{M}
    \Phi_{0}^{(m)}(f)
    \left[\Phi_{1}^{(m)}(f)\right]^{*},
\end{equation}
where $\Phi_{0,1}^{(m)}(f)$ are the Fourier transforms of the phase fluctuations retrieved from the two channels for the $m$th record or time segment. This averaging preserves phase fluctuations that are correlated between the two optical probes, while the statistically independent noise averages towards zero. The residual uncertainty associated with these uncorrelated terms decreases approximately as $1/\sqrt{M}$. Fig.~\ref{fig3}f demonstrates this principle for a carrier at $f_{\mathrm{THz}}=270.594$~GHz, generated by a frequency extender with $N=27$. At low offset frequencies, the cross-spectrum agrees with the two single-channel measurements and with the expected multiplied RF-source phase noise. At larger offset frequencies, where the individual channels reach their respective readout floors, the cross-correlated measurement suppresses the uncorrelated background by approximately 12~dB and extends the useful phase-noise measurement range.

Our results highlight the central advantage of the comb-referenced photonic approach: precise carrier-frequency readout and phase-noise analysis at widely separated mmWave/THz frequencies can be performed with the same receiver hardware, using the comb as the frequency reference. Furthermore, the phase-noise levels retrieved in Fig.~\ref{fig3}d place the demonstrated receiver in a range directly relevant to mmWave and sub-THz source characterization. Representative integrated oscillators and phase-locked sources in this frequency range report single-sideband phase-noise levels of order $-50$ to $-90$~dBc/Hz at kHz-to-MHz offset frequencies. For example, a 300~GHz InP-HBT PLL reported phase noise of $-54$~dBc/Hz at 1~kHz offset and $-78$~dBc/Hz at 100~kHz offset, a W-band CMOS PLL operating from 78.84 to 84~GHz reported $-85.2$~dBc/Hz at 1~MHz offset, and a D-band SiGe:C HBT VCO covering 114.2--148~GHz reported $-94.4$~dBc/Hz at 1~MHz offset~\cite{seo2011_300ghz_pll,trinh2022_wband_pll,zaben2025_dband_vco}. These values overlap the range investigated here, indicating that the comb-referenced TFLN receiver can be used not only to recover imposed phase noise, but also to characterize practical VCOs, PLLs and frequency-multiplied local-oscillator chains at mmWave and sub-THz carrier frequencies.

\tocless\subsection{Mutual coherence and phase-noise correlation of mmWave frequency comb lines}

Finally, we show that combining the two signal analysis techniques discussed in Fig.~\ref{fig2} and \ref{fig3} opens up the possibility to synchronously measure multiple mmWave/terahertz signals distributed across more than 100~GHz and quantify their mutual coherence. In conventional electronic measurements, accessing the relative phase noise of such widely separated carriers would require multiple phase-coherent down-conversion paths with frequency-specific local oscillators, making the measurement complex and difficult to calibrate. Here, the TFLN receiver captures the entire set in a single acquisition, mapping each carrier to a distinct aliased intermediate-frequency tone.

We apply our technique to study the phase correlations of the comb lines emitted by a single mmWave frequency comb. A microwave source at frequency $f_{\mathrm{MW}}$ drives a frequency extender, generating harmonics $f_m=m f_{\mathrm{MW}}$ in the WR\,2.8 band. Since these harmonics originate from the same microwave reference, their phase fluctuations are expected to be mutually correlated, with the phase-noise level increasing according to the usual multiplication scaling. 

\begin{figure}[h!]
    \centering
    \includegraphics[width=18cm]{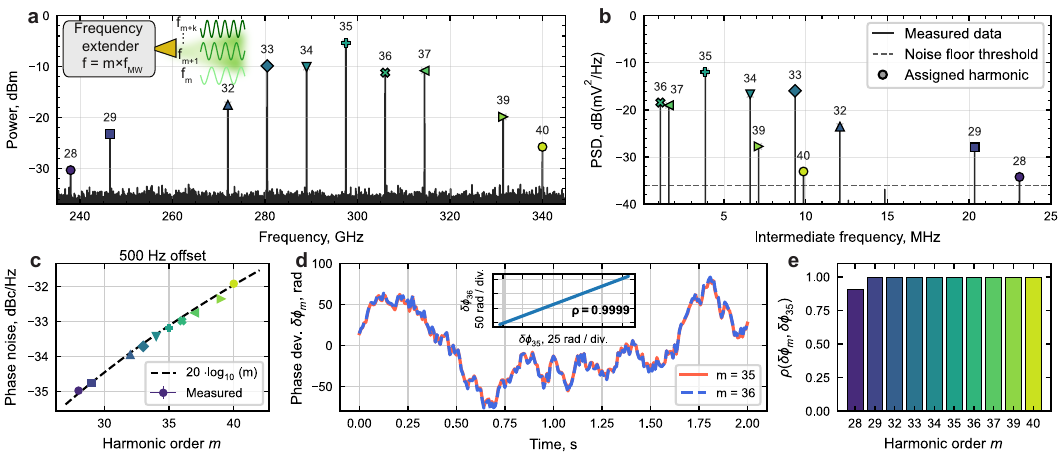}
    \renewcommand{\baselinestretch}{1} 
    \caption{\textbf{Mutual coherence and phase-noise correlation of a frequency comb in the WR\,2.8 band.}
    \textbf{(a)} Measured mmWave/THz spectrum generated by a frequency extender driven at $f_{\mathrm{MW}}$=8.5 GHz, producing harmonics $f_m=m f_{\mathrm{MW}}$ in the WR\,2.8 band. Assigned harmonic numbers are indicated above the peaks.
    \textbf{(b)} Corresponding aliased photodetected spectrum, where each harmonic is mapped to a distinct intermediate-frequency tone. Harmonic assignment is obtained from the known coarse carrier frequencies and the comb-referenced aliasing relation.
    \textbf{(c)} Measured single-sideband phase noise at 500~Hz offset as a function of harmonic order $m$, compared with the expected $20\log_{10}(m)$ scaling.
    \textbf{(d)} Example phase-deviation temporal traces for harmonics $m=35$ and $m=36$. Inset: phase-phase correlation showing near-linear dependence.
    \textbf{(e)} Correaltion coefficient $\rho(\delta \phi_{35}, \delta \phi_m)$ between the phase deviation of harmonic $m=35$ and the other detected harmonics.}
    \renewcommand{\baselinestretch}{1.5} 
    \label{fig4}  
\end{figure}

Fig.~\ref{fig4}a shows the WR\,2.8 spectrum reconstructed using the heterodyne scanning method described in Fig.~\ref{fig2}. The frequency extender was driven by a microwave source at $f_{\mathrm{MW}}$=8.5 GHz and $\approx$-64 dBc/Hz phase noise at 500 Hz frequency offset, producing harmonics from $m=28$ to $m=40$. The assigned harmonic numbers are indicated above the measured peaks. 

The same harmonic set is then detected with the phase-resolved readout (Fig.~\ref{fig3}a-d), producing the aliased photodetected spectrum shown in Fig.~\ref{fig4}b. Each harmonic maps to a distinct intermediate-frequency tone according to $f_{\mathrm{IF},m}=|m f_{\mathrm{MW}}-k_m f_{\mathrm{rep}}|$, where $k_m = \frac{m f_\mathrm{MW}}{f_\mathrm{rep}}$ allowing the tones to be individually selected and digitally demodulated to recover their time-dependent phase deviations $\delta\phi_m(t)$. Fig.~\ref{fig4}~c shows the individual phase noise of all measured harmonics at a 500 Hz offset as a function of harmonic order $m$, together with the expected $20\log_{10}(m)$ scaling. Since all harmonics originate from the same microwave drive, their absolute phase-noise amplitude scales with harmonic number, as expected. This confirms that the detected phase fluctuations are dominated by phase noise inherited from the common microwave reference and transferred through the multiplication chain.

Furthermore, the synchronous phase-resolved readout allows direct comparison of the temporal phase trajectories of different harmonics. If two carriers are mutually coherent, their phase deviations should remain strongly correlated, apart from the absolute phase-noise amplitude scaling with harmonic number. Conversely, independently generated or decorrelated carriers would produce weakly correlated phase traces. To quantify the degree of correlation, we computed the correlation coefficient between harmonics $m$ and $n$ as follows:
\begin{equation}
    \rho(\delta \phi_m, \delta \phi_n) = \frac{\mathrm{cov}(\delta \phi_m, \delta \phi_n)}{\sqrt{\mathrm{var}(\delta \phi_m )\mathrm{var}(\delta \phi_n)}} 
\end{equation}

As an example, Fig.~\ref{fig4}d compares the recovered phase deviations for harmonics $m=35$ and $m=36$. The two traces follow the same temporal fluctuations, and the phase-phase plot in the inset shows an almost linear relation with $\rho\simeq1$. We quantify this behavior across the harmonic set by performing linear regression between $\delta\phi_{35}(t)$ and the phase deviation of each detected harmonic.  The resulting coefficients of determination are shown in Fig.~\ref{fig4}e. Most harmonics exhibit $\rho$ values close to unity, demonstrating strong mutual coherence across carriers separated by tens of gigahertz. Additional details on the data treatment are provided in the supplementary information~\ref{si_note4}.

These measurements demonstrate that our TFLN receiver can not only measure the phase noise of individual mmWave/THz carriers, but also quantify mutual coherence across a broadband multi-harmonic spectrum in a single acquisition. This capability is relevant for characterizing frequency-multiplied sources, broadband comb-like emitters, multi-carrier communication signals, and phase-coherent mmWave/THz metrology systems.

\tocless\section{Conclusions and outlooks}\label{sec_conclusions}

In this work, we demonstrated an antenna-coupled TFLN electro-optic receiver for broadband mmWave/THz spectrum, phase-noise, and mutual-coherence analysis. We characterized mmWave/THz carriers from 80 to 380~GHz across the WR,9.0 and WR,2.8 bands, achieving optical scan speeds up to 25~THz/s, carrier frequency error readouts comparable to the resolution bandwidth, and a relatively flat displayed average noise level below -104~dBm/Hz. Using mode-locked fs laser sampling, we further measured phase noise at 90.225 and 270.625~GHz without a tunable electronic local oscillator; dual-wavelength cross-correlation suppressed uncorrelated photodetection and digitization noise, lowering the measurement floor by more than 12~dB. Finally, simultaneous phase-resolved acquisition quantified the phase-noise scaling and mutual coherence of multiple harmonics separated by tens to more than 100~GHz across the WR2.8 band, avoiding the multiple phase-coherent down-conversion paths required in conventional electronic implementations.

These results position the electro-optic PIC analyzer as a complementary architecture to both electronic spectrum analyzers with frequency extenders and photomixer-based photonic spectrum analyzers. As summarized in Table~\ref{tab:architecture_comparison}, the present implementation combines broad single-setup frequency coverage, chip-scale electro-optic reception, optical-frequency-referenced carrier readout, direct phase- and intensity-noise retrieval, and simultaneous mutual-coherence analysis of widely separated carriers, but does not yet match the best-reported DANL of mature electronic or photomixer systems.

\begin{table*}[h!]
\centering
\begin{threeparttable}
\caption{\textbf{Comparison of millimeter-wave and terahertz spectrum-analysis architectures.}}
\label{tab:architecture_comparison}

\footnotesize
\renewcommand{\arraystretch}{1.18}
\setlength{\tabcolsep}{4pt}

\begin{tabularx}{\textwidth}{p{0.18\textwidth}YYY}
\toprule
\textbf{Metric} &
\textbf{Electronic ESA + extenders\tnote{a}} &
\textbf{Optoelectronic photomixer PSA\tnote{b}} &
\textbf{This work: electro-optic PIC} \\
\midrule

\textbf{Single-setup frequency span} &
Native RF/mmWave ESAs reach up to $\sim 100$ GHz; THz operation requires one waveguide band per extender and is scalable by changing extender modules &
Broad single-setup coverage reported from $<50$ GHz to 6.5 THz &
Broad single-setup coverage demonstrated from 80 to 380 GHz \\

\textbf{DANL / noise floor} &
Representative values: $-139$ dBm/Hz at 100 GHz, $-115$ dBm/Hz at 1 THz and $-107$ dBm/Hz at 1.5 THz &
Reported $-145.6$ dBm/Hz at 100 GHz, $-134$ dBm/Hz at 1 THz and extrapolated $-105.2$ dBm/Hz at 6.5 THz &
Measured DANL of $-105$ dBm/Hz \\

\textbf{Scan speed in spectrum-analyzer mode} &
Up to $\sim 3~\mathrm{THz\,s^{-1}}$ for RF/mmWave real-time ESAs; 
broad THz coverage with external extenders remains waveguide-band specific &
Up to $\sim 1~\mathrm{THz\,s^{-1}}$ using a swept photonic local oscillator &
0.05-25~$\mathrm{THz\,s^{-1}}$ optical-frequency sweep with time-domain EO readout\tnote{b} \\

\textbf{Phase/intensity-noise analysis} &
Mature standard capability within the selected analysis band &
Frequency stability and linewidth analysis demonstrated; broadband phase- and intensity-noise analysis of arbitrary external carriers not demonstrated &
Direct phase- and intensity-noise retrieval from the time-domain optical readout \\

\textbf{Instantaneous signal-analysis bandwidth} &
High; GHz-class ESA bandwidth and extender IF bandwidths up to tens of GHz &
Local electrical IF around the photonic LO, limited by the readout chain; reported from kHz-scale to tens of MHz &
Alias-free IF window set by $f_\mathrm{rep}/2$; 50 MHz demonstrated in this work. 
Noise-analysis bandwidth is set by the photodetector/digitizer readout. \\

\textbf{Mutual-coherence analysis} &
Possible within a common captured IF window; broadband cross-band analysis requires synchronized multi-channel hardware &
Not demonstrated as simultaneous broadband mutual-coherence analysis of arbitrarily separated carriers &
Demonstrated using a common optical readout of multiple carriers separated by >100 GHz \\

\textbf{Integrated receiver front end} &
Discrete RF/mmWave/THz hardware chain &
Chip-scale photomixer/PCA receiver; external photonic LO and electronics. Broader photomixers/photodetectors integration with lasers and optical amplifiers\tnote{c} &
Chip-scale EO-PIC receiver; external laser, PD, reference and electronics. Broader TFLN integration with lasers and high-speed photodiodes has been demonstrated\tnote{d} \\

\bottomrule
\end{tabularx}

\begin{tablenotes}
\footnotesize
\item[a] Electronic ESA/extender specifications and representative values from Keysight, Aaronia and VDI.~\cite{keysight_n9042b, aaronia_spectran_v6, vdi_sax}
\item[b] Optoelectronic photomixer PSA values from Ref.~\cite{krause2025natcom}
\item[c] Refs.~\cite{zhao2026terahertz}
\item[d] Refs.~\cite{op2021iii, zhang2022ol_iiiv_ln, shams2022electrically, li2025heterogeneously, zheng2026heterogeneously}
\end{tablenotes}

\end{threeparttable}
\end{table*}
Heterogeneous integration can directly improve the demonstrated analyzer by reducing optical loss, noise, and alignment sensitivity. In the present system, the TFLN receiver is connected to off-chip lasers, EDFAs, and photodiodes, and its sensitivity is partially limited by chip-coupling loss and residual optical noise at the detector. Co-packaging these components would eliminate multiple fiber interfaces, increase the optical power delivered to the chip, and enable on-chip optical amplification and photodetection. For example, integrated amplifiers could compensate for chip insertion loss, and integrated high-speed photodiodes could replace external fiber-coupled detection. These changes would directly improve the detected SNR and reduce the DANL. Recent progress in heterogeneous III--V-on-LNOI light sources, amplifiers, and photodetectors \cite{op2021iii, zhang2022ol_iiiv_ln, shams2022electrically, li2025heterogeneously, zheng2026heterogeneously}, transfer-print integration of TFLN devices onto silicon photonics \cite{tan2024acsp_tfln_mtp}, and wafer-scale LN integration on low-loss passive platforms \cite{he2019natphotonics, churaev2023natcom_lnsin} provides a route toward such a compact and stable single-module mmWave/THz analyzer.
In parallel, the operating frequency range can be extended beyond the bands explored here by improving the THz interface, including the antenna, transmission line, and packaging. Because the Pockels effect is intrinsically broadband, the main scaling challenge is efficient coupling of higher-frequency fields to the optical mode with low loss and sufficient electro-optic overlap. Recent progress on photonics-integrated THz transmission lines in lithium niobate \cite{lampert2025natcom, herter2025natcom_thzdet}, together with the broadband antenna--transmission-line response predicted in our simulations, provides a concrete roadmap for extending the receiver up to 2.5 THz operation. Together, broadband TFLN electro-optic transduction, comb-referenced sampling, and heterogeneous photonic integration suggest a realistic path toward compact mmWave/THz spectrum, phase-noise, and coherence analyzers with expanded frequency coverage, reduced calibration burden, and improved system-level robustness.

\tocless\section{Methods}
\tocless\subsection{Fabrication}\label{fabrication}
The devices were fabricated on an X-cut thin-film lithium niobate on insulator (LNOI) wafer, consisting of a $600$-nm-thick thin-film lithium niobate layer on a $4.7~\mu\mathrm{m}$ $\mathrm{SiO_2}$ buried oxide layer supported by a high-resistivity silicon substrate. Photonic structures were defined by electron-beam lithography using a FOX16 hard mask and developed in a 25\% TMAH solution. The pattern was transferred into the lithium niobate layer by ion-beam etching operated at $150$~W for $15$~min with a $10^\circ$ incidence angle, resulting in rib waveguides with an etch depth of approximately $350$~nm and sidewall angles of about $75^\circ$. Lithium-niobate redeposition was removed in a heated solution of 40\% KOH and 30\% $\mathrm{H_2O_2}$ (3:1 by volume) at $85^\circ$C. The coplanar transmission line and on-chip antenna were fabricated by a lift-off process using an MMA/PMMA bilayer resist. After electron-beam exposure and development, a $5$-nm Ti adhesion layer and a $300$-nm Au layer were sequentially deposited by electron-beam evaporation, followed by lift-off in acetone, rinsing in IPA, and drying under nitrogen flow.
\tocless\subsection{Setup details for fast broadband spectrum analysis with an antenna-coupled TFLN PIC}\label{setupdetails_fast}

The experimental setup is schematically shown in Fig.~\ref{fig2}a of the main text. The fixed-wavelength pump laser was an Aerodiode 1550LD-6-0-0 DFB laser, operating near 1550~nm with an output power of 10~mW and an instantaneous linewidth of 40~kHz. The laser was driven by a Koheron CTL200 controller and amplified by an erbium-doped fiber amplifier (Nuphoton stretched MSA PM 1 W EDFA, model EDFA-C0-PM-SMR-30-20-FCA) up to 400~mW before being coupled to the TFLN PIC. For wavelength-swept measurements, we used either a Keysight N7776C external-cavity laser (ECL), operated at tuning speeds from 600~GHz/s to 25~THz/s, or a thermally tuned YouOpto YTFCDD02BK1IH-1550 DFB laser, with 40~mW output power and 69~kHz instantaneous linewidth, driven by a second Koheron CTL200 controller. In the latter case, the wavelength was thermally swept at approximately 50~GHz/s.

The mmWave/THz radiation was generated using an Anritsu MG362x1A RF/microwave signal generator, which provided a 9--14~GHz input to Virginia Diodes WR\,9.0 and WR\,2.8 signal-generator extension modules. The modules performed $9\times$ or $27\times$ frequency multiplication, delivering up to 100~mW in the WR\,9.0 band and up to 5~mW in the WR\,2.8 band. The generated radiation was emitted into free space using horn antennas with gains of 21~dBi and 26~dBi in the WR\,9.0 and WR\,2.8 bands, respectively.

The optical sweep was calibrated using a Menlo C-Fiber 780 femtosecond laser, operated at its 1560~nm output with a repetition rate of $f_{\mathrm{rep}}=99.967$~MHz. A fraction of the swept tunable laser was heterodyned with the comb, and the resulting beat signals were used to reconstruct the instantaneous laser detuning during the scan. The optical heterodyne signals were detected using Koheron PD100-AC photodiodes and digitized with a Spectrum Instrumentation M4i.2442-x8 acquisition card. The digitizer input bandwidth was set to 20~MHz to avoid aliasing. The recorded data were processed using self-developed Python code.

\tocless\subsection{Setup details for comb-referenced TFLN photonic phase-noise analyzer}\label{setupdetails_PN}
The experimental setup is schematically shown in Fig.~\ref{fig3}a of the main text. The mmWave/THz radiation was generated as described in the previous section, using a Keysight E8267D PSG Vector Signal Generator as the low-frequency input to the frequency-multiplier chain. This allowed controlled phase-noise levels to be imposed on the generated carrier by applying programmed Gaussian phase noise to the RF source before multiplication. The imposed phase noise was independently characterized using a Keysight N9030A electrical spectrum analyzer. The optical sampling source and photodetection electronics were based on the same platform used for the broadband spectrum measurements. For phase-noise acquisition, the M4i.2442-x8 digitizer was operated without the 20~MHz input-bandwidth limitation in order to preserve the full bandwidth of the detected aliased signal. The output signal was further amplified using an erbium-doped fiber amplifier. The recorded data were processed using self-developed Python code, as described below.

For the cross-correlation measurements shown in Fig.~\ref{fig3}e-f, two synchronized pulse trains centered at 1550 and 1310~nm were generated using a Levante IR fs OPO and a Coherent Chameleon Discovery NX OPO, respectively. The two wavelength channels were coupled through dedicated grating couplers into separate input waveguides of the TFLN PIC and combined on chip using Y-couplers. Both optical probes therefore traversed the same antenna-coupled electro-optic interaction region and acquired the phase fluctuations of the incident mmWave/THz carrier. After the interaction region, the two wavelength channels were routed to independent output paths. The corresponding grating couplers were designed for the 1550 and 1310~nm wavelength bands, respectively, and thus provided wavelength-selective in- and out-coupling that suppressed cross-coupling between the two optical channels. The two outputs were detected on independent photodiodes and simultaneously acquired on two channels of the M4i.2442-x8 digitizer.
\vspace{1cm}

\textbf{Data Availability}
The data generated in this study will be made available in the Zenodo database prior to publication.

\medskip
\textbf{Code Availability}
The code used to plot the data within this paper will be made available in the Zenodo database prior to publication.

\bibliographystyle{paper}
\bibliography{bibliography}

\medskip
\textbf{Acknowledgments}
A.G. and I.C.B.C. acknowledge funding from the European Union’s Horizon Europe research and innovation programme under project MIRAQLS with grant agreement No. 101070700 and funding from the Swiss National Science Foundation Grant No. №219406. E.M. acknowledges funding under the MSCA-DN grant agreement No. 101169225 – SPARKLE. J.L. acknowledges funding through the Excellence Postdoctoral Fellowship Programme from the Quantum Science Center at EPFL. The chips were fabricated at the Center of MicroNanoTechnology (Cmi) at EPFL.

\medskip
\textbf{Author contributions} A.G. conceived the experiment, designed the photonic devices, developed the measurement and analysis methods, carried out the experiments, analyzed the data, and wrote the manuscript. E. M. designed the chip for the cross-correlation experiment with dual-color pump presented in Fig.3e-f. G.J. and E. M. assisted with the experimental setup and the realization of the heterodyne detection scheme. J.L. fabricated the devices. I.C.B.C. supervised the project and contributed to writing the manuscript. All authors discussed the results and reviewed the manuscript.

\medskip
\textbf{Competing interests}
A.G. and I.C.B.C. are co-inventors on a patent application regarding the phase- and intensity noise retrieval method. The other authors declare no competing interests. 

\medskip

\textbf{Corresponding authors} Correspondence to Aleksei Gaier (aleksei.gaier@epfl.ch) or Ileana-Cristina Benea-Chelmus (cristina.benea@epfl.ch).

\appendix

\renewcommand{\thefigure}{S\arabic{figure}}
\setcounter{figure}{0} 
\setcounter{linenumber}{1}
\setcounter{equation}{0}
\renewcommand{\theequation}{S\arabic{equation}}
\clearpage
\setcounter{page}{1}
\begin{center}
    {\Huge \bfseries Supplementary Information}
\end{center}

\tableofcontents

\section{Photonic integrated circuit operation principles and design} \label{PIC_design}
\begin{figure}[h!]
    \centering
    \includegraphics[width=16cm]{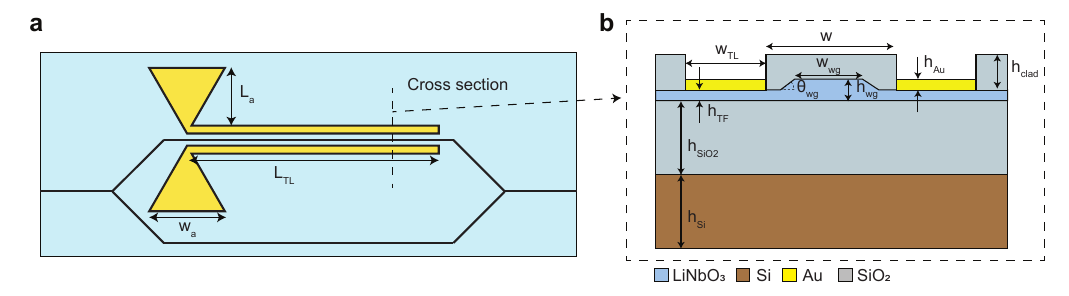}
    \renewcommand{\baselinestretch}{1} 
    \caption{\textbf{Sketch of the device used in this work}}
    \textbf{(a)} Top view
    \textbf{(b)} Cross section
    \renewcommand{\baselinestretch}{1.5} 
    \label{figS1}  
\end{figure}

Light at a wavelength around 1550nm is coupled in and out through grating couplers, with a fiber-to-fiber coupling efficiency of -23~dB (characterized independently on a separate device without a Mach-Zehnder interferometer). The same characterization was performed for the grating at 1310nm, yielding similar performance. After entering the chip, the optical field is routed into a Mach-Zehnder interferometer (MZI) operated at the quadrature point. One arm of the interferometer is loaded with a coplanar stripline (CPS) transmission line that guides the mmWave/THz field coupled from free space via an on-chip bow-tie antenna (the sketch off the device is shown in Fig.\ref{figS1}), and induces an optical phase shift via the Pockels effect in thin-film lithium niobate. 
\begin{table} [h!]
    \centering
    \begin{tabular}{ccc}
         thickness of the high resistivity silicon substrate & $h_{Si}$ & 500 ${\mu }$m\\ 
         thickness of the silicon oxide layer & $h_{SiO_2}$ & 4.7 ${\mu }$m \\
         thickness of thin-film lithium niobate layer & $h_{TF}$  & 300 nm \\
         thickness of the gold layer & $h_{Au}$  & 300 nm \\
         height of the lithium niobate waveguide & $h_{wg}$  & 600 nm \\
         thickness of the silicon oxide cladding & $h_{clad}$  & 1 $\mu $m \\
         width of the lithium niobate waveguide & $w_{wg}$  & 1.5 $\mu $m \\
         etching angle of the lithium niobate waveguide & $\theta_{wg}$ & $\ang{75}$
         \\
         distance between the electrodes & $w$  & 4 ${\mu }$m \\
         width of the antenna & $w_{a}$  & 225 $\mu $m \\
         width of the the transmission line & $w_\mathrm{TL}$  & 2 $\mu $m \\
         length of the transmission line & $L_\mathrm{TL}$  & 3 mm \\
         length of the bow-tie antenna & $L_{a}$  & 225 $\mu $m \\
    \end{tabular}
    \caption{Dimensions of the device used in this work}
    \label{tab:dimensions}
\end{table}

The antenna-coupled TFLN device can be operated in two optical readout configurations. In the broadband spectrum measurements, the device is used as a phase modulator (PM): the mmWave/THz field directly generates optical sidebands on the transmitted optical carrier, which are subsequently detected by heterodyne readout. In the phase-resolved measurements, the same electro-optic phase shift is read out using a Mach--Zehnder interferometer (MZI) biased near quadrature, which converts the pulse-to-pulse phase modulation into intensity modulation for direct photodetection.

In both cases, the mmWave/THz-induced phase shift in the modulated arm is~\cite{lampert2025natcom, gaier2025wireless}

\begin{equation}
    \Delta \phi_m(t)= \frac{\chi^{(2)}}{c_{0}n_o} \, \omega_{o} \Gamma_{\mathrm{eo}} L_{\mathrm{TL}} PM \cdot E_{\Omega}(t),
    \label{mod-eff-eq}
\end{equation}

where $\chi^{(2)} = \chi^{(2)}_{333}$ is the largest second-order nonlinear susceptibility tensor component of lithium niobate, $n_o$ is the effective refractive index of the optical mode, $\omega_o$ is the optical angular frequency, $\Gamma_{\mathrm{eo}}$ is the optical--mmWave/THz overlap integral, and $L_{\mathrm{TL}}$ is the interaction length. The phase-matching function

\begin{equation}
    PM =
    \frac{e^{i \Delta \tilde{k} L_{\mathrm{TL}}} - 1}
    {\Delta \tilde{k} L_{\mathrm{TL}}}
\end{equation}

accounts for both propagation phase mismatch and mmWave/THz loss along the transmission line, with

\begin{equation}
    \Delta \tilde{k}
    =
    \frac{\Omega}{c_0}(n_m-n_g)
    +
    i\frac{\alpha_{m}}{2}.
\end{equation}

Here, $n_g$ is the optical group index, $n_m$ is the effective index of the mmWave/THz mode, and $\alpha_{m}$ is the mmWave/THz power attenuation coefficient.

For a sinusoidal mmWave/THz field,

\begin{equation}
    E_{\Omega}(t)=E_m\cos(\Omega t),
\end{equation}

the phase shift can be written as

\begin{equation}
    \Delta\phi_m(t)=\beta \cos(\Omega t),
\end{equation}

where the peak phase-modulation amplitude is

\begin{equation}
    \beta =
    \frac{\chi^{(2)}}{c_{0}n_o}
    \omega_o \Gamma_{\mathrm{eo}} L_{\mathrm{TL}} PM \cdot E_m .
    \label{eq:beta_definition}
\end{equation}

The field amplitude $E_m$ is related to the coupled mmWave/THz power as

\begin{equation}
    E_m =
    \sqrt{
    \frac{2\eta_{\mathrm{coupl}}P_m}
    {n_m c_0 \varepsilon_0 S_m}
    },
\end{equation}

where $\eta_{\mathrm{coupl}}$ is the fraction of the incident free-space mmWave/THz power $P_m$ coupled into the transmission-line mode, and $S_m$ is the effective mmWave/THz mode area.

For the PM readout, the output optical field is

\begin{equation}
    E_{\mathrm{PM}}(t)
    =
    E_{\mathrm{0}} e^{i\beta \cos(\Omega t)}.
\end{equation}

In the small-signal limit, $\beta\ll 1$, the first-order sideband-to-carrier power ratio at each sideband is

\begin{equation}
    \left.
    \frac{P_{\pm1}}{P_0}
    \right|_{\mathrm{PM}}
    \approx
    \left(\frac{\beta}{2}\right)^2 .
    \label{eq:pm_sideband_ratio}
\end{equation}
where $P_0$ is the power of the optical carrier, and $P_{\pm1}$ are the powers of the first-order optical sidebands.
For the MZI readout, the phase modulation is applied in one interferometer arm and converted to intensity modulation at the output. For an MZI biased at quadrature, the output intensity can be written as

\begin{equation}
    I_{\mathrm{MZI}}(t)
    =
    \frac{I_0}{2}
    \left[
    1 - \sin\left(\Delta \phi_m(t)\right)
    \right],
\end{equation}

which reduces in the small-signal limit to

\begin{equation}
    I_{\mathrm{MZI}}(t)
    \approx
    \frac{I_0}{2}
    \left[
    1 - \beta\cos(\Omega t)
    \right].
\end{equation}

Thus, the MZI provides a linear conversion from the mmWave/THz-induced optical phase shift to intensity modulation. However, for the same input optical power and the same phase-modulation amplitude $\beta$, the first-order optical sidebands at one MZI output are reduced compared with the pure PM case. In the small-signal limit, the sideband power at each sideband scales as

\begin{equation}
    \left.
    \frac{P_{\pm1}}{P_{\mathrm{0}}}
    \right|_{\mathrm{MZI}}
    \approx
    \left(\frac{\beta}{4}\right)^2 ,
    \label{eq:mzi_sideband_absolute}
\end{equation}

whereas for the PM readout

\begin{equation}
    \left.
    \frac{P_{\pm1}}{P_{\mathrm{0}}}
    \right|_{\mathrm{PM}}
    \approx
    \left(\frac{\beta}{2}\right)^2 .
\end{equation}

Therefore, for the same $\beta$ and input optical power, the PM readout provides a factor of four higher sideband power, corresponding to a 6~dB larger response. The broadband spectrum measurements in the main text were therefore performed in the PM configuration to maximize sensitivity. The MZI configuration can also be used for broadband spectrum analysis, but with this additional 6~dB response penalty. Its advantage is phase-resolved readout: when the MZI is probed with a femtosecond pulse train, each pulse samples the instantaneous local mmWave/THz electric field, and the quadrature transfer function converts the sampled Pockels-induced phase shift into pulse-to-pulse intensity modulation. The detected intensity fluctuations therefore retain the temporal phase and amplitude information of the carrier.

Eq.~\ref{mod-eff-eq} shows that, for a given free-space mmWave/THz power, the modulation depth is determined by two main design blocks of the device. First, the transmission-line design determines the effective mode area $S_m$, the effective refractive index $n_m$, the overlap integral $\Gamma_{\mathrm{eo}}$, and the propagation loss through $\alpha_m$, and therefore sets the strength and effective length of the electro-optic interaction. Second, the antenna design determines the coupling efficiency $\eta_{\mathrm{coupl}}$, i.e. the fraction of the incident free-space mmWave/THz power that is transferred to the transmission line and made available for modulation.

These considerations define the main trade-offs in the device design. On the transmission-line side, the goal is to tightly confine the mmWave/THz mode in the vicinity of the optical waveguide in order to minimize $S_\Omega$ and enhance the electro-optic interaction, while simultaneously maintaining low propagation loss and good phase matching to the optical mode. Low loss and favorable phase matching are both essential for preserving a long effective interaction length, since excessive attenuation or accumulated phase mismatch reduces the contribution of distant sections of the line to the total modulation. However, placing the metal electrodes too close to the optical waveguide can increase the overlap between the optical mode and the metal and thereby introduce additional optical absorption loss. Our design, therefore, aims to balance strong mmWave/THz field confinement with low RF loss, low optical loss, and limited phase mismatch in order to maximize the effective interaction over the full transmission-line length.
\begin{figure}[h!]
    \centering
    \includegraphics[width=\linewidth]{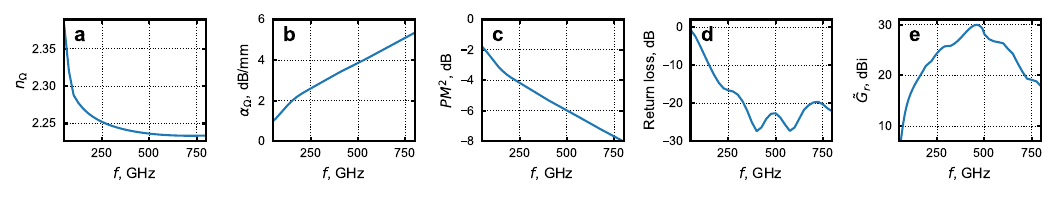}
    \renewcommand{\baselinestretch}{1} 
    \caption{\textbf{Simulated frequency dependence of the electrical and electro-optic parameters used in the model:}
    \textbf{(a)} mmWave/THz effective index $n_m$,
    \textbf{(b)} propagation loss $\alpha_m$,
    \textbf{(c)} phase-matching function $PM$,
    \textbf{(d)} return loss, and
    \textbf{(e)} realized antenna gain $\tilde{G}_r$.}

    \renewcommand{\baselinestretch}{1.5} 
    \label{figS2}  
\end{figure}

On the antenna side, the objective is to maximize the fraction of free-space power delivered to the transmission line over the target frequency range. From the Friis transmission equation \cite{friis1946},
\begin{equation}
    \eta_{\mathrm{coupl}} = G_{\mathrm{t}} \Tilde{G}_{\mathrm{r}} \left(\frac{\lambda}{4\pi R}\right)^2,
    \label{eq:friis}
\end{equation}
where $G_{\mathrm{t}}$ is the transmitter antenna gain, $\Tilde{G}_{\mathrm{r}}$ is the receiver realized antenna gain, $\lambda$ is the free-space wavelength, and $R$ is the separation between the antennas. This expression shows that the received power scales with the realized gain of the receiving antenna, such that higher realized gain directly increases the power available at the chip input. In our case, this must be achieved over a broad bandwidth, since the device is designed to operate across a wide mmWave/THz frequency range. The realized antenna gain, therefore, becomes a key figure of merit, as it combines the effects of directivity, radiation efficiency, and impedance matching to the transmission line.

The main design parameters of the device are summarized in Tab.~\ref{tab:dimensions}. On the transmission-line side, the simulated parameters include the mmWave/THz effective index $n_{\Omega}$ (Fig.~\ref{figS2}a), propagation loss $\alpha_{\Omega}$ (Fig.~\ref{figS2}b), effective mode area $S_{m}\approx 15.4~\mu\mathrm{m}^2$, and optical-mmWave/THz overlap factor $\Gamma_{\mathrm{eo}}\approx 0.6$. These parameters yield a flat phase-matching function $PM$ over the 50-800~GHz band (Fig.~\ref{figS2}c). On the antenna side, the relevant parameters are the impedance matching to the CPS transmission line and the realized gain $\Tilde{G}_{\mathrm{r}}$ (Fig.~\ref{figS2}e), which determine how efficiently free-space mmWave/THz power is transferred to the transmission line. The impedance matching is characterized by the return loss $20\log_{10}\left| \frac{Z_{\mathrm{ant}} - Z_{\mathrm{TL}}}{Z_{\mathrm{ant}} + Z_{\mathrm{TL}}} \right|$, which remains below $-10$~dB above 125~GHz (Fig.~\ref{figS2}d).

\section{Additional details on fast broadband spectrum analysis with an antenna-coupled TFLN PIC}\label{SInote2}

\subsection{Comb-referenced calibration of the laser sweep}

\begin{figure}[h!]
    \centering
    \includegraphics[width=\linewidth]{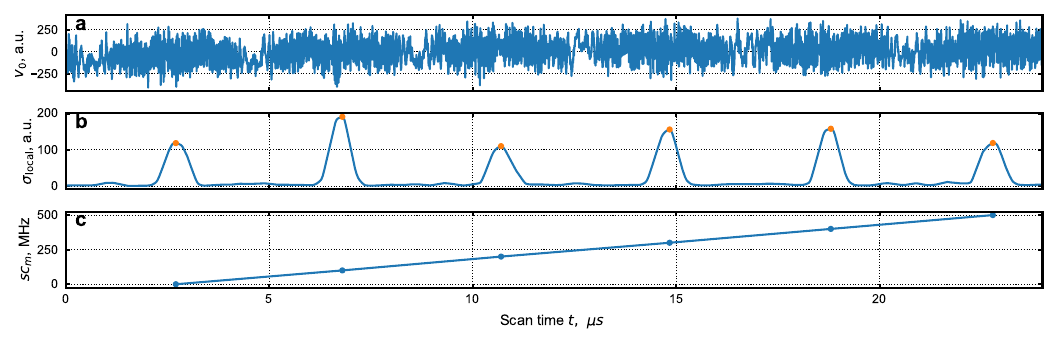}
    \caption{
    \textbf{Comb-referenced calibration of the swept laser frequency.}
    (a) Representative segment of the heterodyne signal between the swept laser and the optical frequency comb.
    (b) Local variance $\sigma_{\mathrm{local}}$ trace used for calibration-tick detection, with identified peaks marked.
    (c) Recovered spectral coordinate $(t_{\mathrm{sc}, m}, sc_m)$.
    }
    \label{fig:comb_calibration}
\end{figure}

To calibrate the instantaneous optical frequency during the laser sweep, we heterodyned the tunable laser with an optical frequency comb and extracted the temporal positions at which the swept laser crossed successive comb teeth. This procedure provided a time-to-frequency calibration ruler for subsequent processing of the heterodyne measurements. The main steps of the calibration procedure are illustrated in Fig.~\ref{fig:comb_calibration}.

Let the comb repetition rate be $f_{\mathrm{rep}}$. As the tunable laser is swept, its beat note with the nearest comb tooth evolves continuously and passes through characteristic frequencies that recur once per crossed tooth. The detected time instants therefore mark equally spaced optical-frequency intervals and define a set of calibration ticks.

More specifically, the recorded comb-laser beat signal $v_0(t)$, sampled at $f_{\mathrm{rep}}$, is first mixed numerically with a cosine sequence,
\begin{equation}
v_{0,1}[n] = v_0[n]\cos(\pi n h),
\end{equation}
where $h$ is the selected harmonic index: $h=1$ corresponds to calibration ticks spaced by $f_{\mathrm{rep}}$, whereas $h=1/2$ corresponds to ticks spaced by $f_{\mathrm{rep}}/2$. The choice of $h$ depends on the tuning speed of the swept laser. At low tuning speeds, $h=1/2$ is preferred because it increases the density of calibration ticks and therefore improves the sampling of the sweep trajectory. At high tuning speeds, $h=1$ is preferred because adjacent ticks corresponding to $f_{\mathrm{rep}}/2$ may become too closely spaced in time to be reliably resolved. The resulting signal is low-pass filtered to isolate the slowly varying envelope associated with the comb-crossing events. The local variance is then evaluated on the filtered trace $\tilde{v}_{0,1}[n]$ as $\sigma_{\mathrm{local}}[n] = \frac{1}{2N+1}\sum_{k=-N}^{N}\tilde{v}_{0,1}^{,2}[n+k]$, where $2N+1$ is the length of the local evaluation window and a zero local mean is assumed. Local maxima of $\sigma_{\mathrm{local}}[n]$ are used to identify the time instants corresponding to the comb crossings. Representative examples of the raw beat signal and the processed trace used for peak detection at a laser tuning speed of $25~\mathrm{THz/s}$ and $h=1$ are shown in Fig.~\ref{fig:comb_calibration}a,b.

The extracted peak positions $p_m$ define the calibration times
\begin{equation}
    t_{\mathrm{sc},m} = \frac{p_m}{f_{\mathrm{rep}}},
\end{equation}
Since consecutive ticks correspond to equally spaced comb intervals, they are assigned a linear spectral coordinate
\begin{equation}
    sc_{m} = f_{\mathrm{rep}}\, m\, h,
\end{equation}
up to an arbitrary offset. The pair $(t_{\mathrm{sc}, m}, sc_m)$ is illustrated in Fig.~\ref{fig:comb_calibration}c.
Overall, this comb-referenced calibration converts the laser sweep's nonuniform time axis into a calibrated spectral coordinate directly used in the heterodyne signal reconstruction described below.

\subsection{Time–frequency trajectory-based signal reconstruction}\label{si-resolution}

\begin{figure}[h!]
    \centering
     \includegraphics[width=\linewidth]{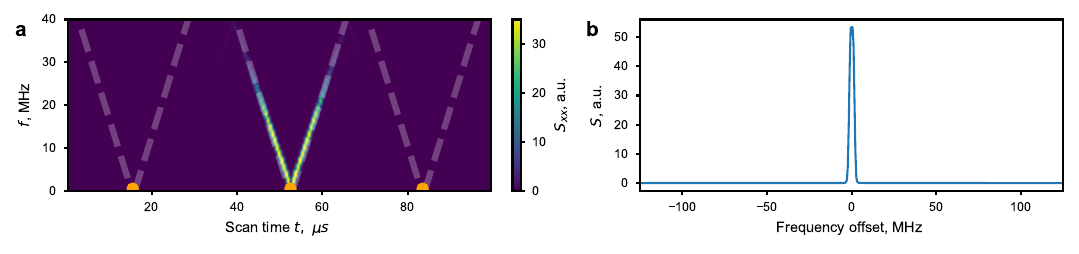}
    \caption{\textbf{
    Trajectory-guided spectral stacking for heterodyne reconstruction.}
    (a) Representative spectrogram $S_{xx}(f,t)$ showing a V-shaped heterodyne feature. The white dashed lines indicate the V-shaped trajectories for the reducer. Points indicate the moments of time $t$ to which these trajectories belong.
    (b) Reconstructed one-dimensional trace $S$ obtained after stacking along the predicted trajectories.
    }
    \label{fig:vradon}
\end{figure}

After establishing the sweep calibration $(t_{\mathrm{sc}, m}, sc_m)$, the heterodyne signal is processed in the time-frequency domain to recover weak beat components. A representative example of the processing workflow is shown in Fig.~\ref{fig:vradon}.

The measured signal $v(t)$ is divided into segments, and for each segment a short-time Fourier transform is computed using a spectrogram representation,
\begin{equation}
    S_{xx}(f,t).
\end{equation}
Because the laser sweep is generally nonlinear, the mapping between laboratory time $t$ and spectral coordinate is obtained by interpolating the calibration times,
\begin{equation}
    sc(t) = \mathcal{I}\!\left(t_{\mathrm{sc}, m}, sc_m\right),
\end{equation}
where $\mathcal{I}$ denotes interpolation. This function provides the sweep laser frequency as a function of scan time.

In an example spectrogram, the heterodyne feature appears as a V-shaped trace in the $(f,t)$ plane, as shown in Fig.~\ref{fig:vradon}a. This shape arises because the beat frequency is given by the absolute difference between the swept laser frequency and the unknown signal frequency. As the swept laser approaches the signal frequency, the beat frequency decreases toward zero, and after the crossing point it increases again, giving rise to two symmetric branches.

To extract the beating signal, we sum up the spectrogram intensity along the two branches of the V-shaped trajectory. For each frequency offset $\Delta f_j = j\,df$, the corresponding temporal offset is denoted by $\Delta k_j(t)$ and is expressed in units of spectrogram frames. Here, $\Delta t$ denotes the time spacing between adjacent spectrogram frames. The offsets $\Delta k_j(t)$ are determined from the calibrated sweep coordinate $sc(t)$ so that the stacking follows the measured sweep dynamics. The values collected along all admissible trajectories are then combined using a robust reducer, typically the median:
\begin{equation}
    S(t) =
    \operatorname{median}_{j,\pm}
    \left\{
    S_{xx}\!\left(f_j,\ t \pm \Delta k_j(t)\Delta t\right)
    \right\}.
\end{equation}
Alternative reducers, such as the mean or max, can also be used, but the median was found to be more robust against outliers and spurious spectrogram features.

Conceptually, this procedure performs a trajectory-based accumulation of spectrogram energy, producing a reconstructed trace $S(t)$ on the laboratory time axis. Because the calibrated sweep coordinate $sc(t)$ is interpolated on the same time grid, this trace is directly remapped as $S(sc)$, as shown in Fig.~\ref{fig:vradon}b. The dominant beating peak is then identified in the reconstructed trace, and the frequency axis is shifted relative to this peak as $f = sc - sc_{\mathrm{peak}}$.

The effective resolution bandwidth of the reconstructed spectrum is determined by the time--frequency resolution of the spectrogram used to resolve the V-shaped heterodyne trajectory. For a sampling rate $f_s$ and an spectrogram window length $N$, the frequency resolution scales as
\begin{equation}
    \mathrm{RBW} \sim \frac{f_s}{N}.
\end{equation}
At the same time, the window must be short enough that the swept beat note does not move by more than approximately one frequency bin during the window. For a local sweep rate $v_{\mathrm{scan}}=\left|dsc/dt\right|$, this condition gives
\begin{equation}
    v_{\mathrm{scan}}\frac{N}{f_s} \sim \frac{f_s}{N},
\end{equation}
and therefore
\begin{equation}
    N \sim \frac{f_s}{\sqrt{v_{\mathrm{scan}}}},
    \qquad
    \mathrm{RBW} \sim \sqrt{v_{\mathrm{scan}}}.
\end{equation}
Thus, faster optical scans increase the reconstructed RBW as the square root of the sweep rate, while slower scans improve spectral resolution. For example, at the maximum scan speed used in this work, $v_{\mathrm{scan}}=25~\mathrm{THz/s}$, this Fourier-limited estimate gives $\mathrm{RBW}\sim5~\mathrm{MHz}$. Experimentally, we measured an RBW of $7.4\pm0.2~\mathrm{MHz}$, with the difference attributed to the numerical prefactor set by the spectrogram window. This scan-speed--RBW trade-off also affects the accumulated signal-to-noise ratio, motivating the noise and SNR analysis in the following section.

\subsection{Signal-to-noise ratio model and noise sources}\label{si2_snr}

The beating signal on a photodiode can be written as
\begin{equation}
    V_\mathrm{sig}
    =
    G_\mathrm{TIA} \mathcal{R}
    \sqrt{2 P_\mathrm{LO} P_{\pm1}} =  G_\mathrm{TIA} \mathcal{R} \beta
    \sqrt{\frac{P_\mathrm{LO} P_\mathrm{0}}{2}} 
\end{equation}
where $V_\mathrm{sig}$ is the RMS voltage amplitude of the heterodyne beat note, $G_\mathrm{TIA}$ is the transimpedance gain in V/A, $\mathcal{R}$ is the photodiode responsivity in A/W, $P_\mathrm{LO}$ is the optical power of the reference laser, $P_{\pm1}$ is the optical power in the electro-optic sideband, and $P_{0}$ is the optical power pumping the chip.

The total noise variance at the photodetector output can be expressed as
\begin{equation}
    V_\mathrm{noise}^2
    =
    S_\mathrm{V, el}^2 B
    +
    2 e \mathcal{R} P_\mathrm{tot} G_\mathrm{TIA}^2 B
    +
    \left(
    \mathcal{R} P_\mathrm{tot} G_\mathrm{TIA}
    \right)^2
    \int_0^B S_\mathrm{rel}(f)\,df ,
\end{equation}\label{noise_model_si2}
where $S_\mathrm{V, el}$ is the power-independent electronic noise variance, $e$ is the elementary charge, $B$ is the detection bandwidth, and
\begin{equation}
    P_\mathrm{tot} = P_\mathrm{LO} + P_\mathrm{0}
\end{equation}
is the total optical power incident on the photodiode. The second term represents optical shot noise. The last term describes relative intensity-like noise, where $S_\mathrm{rel}(f)$ is an effective relative noise spectral density that can include laser relative intensity noise (RIN), EDFA intensity noise, coupling fluctuations, and other power-dependent technical noise sources.

The V-shaped integration algorithm coherently sums the signal over
\[
M \approx \frac{2B}{\mathrm{RBW}}
\]
spectral bins. Assuming uncorrelated noise between bins, the signal remains unchanged while the noise variance is reduced by a factor of \(M\). Consequently,
\begin{equation}
\mathrm{SNR}
=
\frac{V_\mathrm{sig}^2}{V_\mathrm{noise}^2/M}.
\end{equation}

Using the expressions above, this gives
\begin{equation}
    \mathrm{SNR}
    =
    \frac{B}{\mathrm{RBW}} \frac{
    G_\mathrm{TIA}^2 \mathcal{R}^2
    \beta^2 P_\mathrm{LO} P_\mathrm{0}
    }{
    S_\mathrm{V, el}^2 B
    +
    2 e \mathcal{R} P_\mathrm{tot} G_\mathrm{TIA}^2 B
    +
    \left(
    \mathcal{R} P_\mathrm{tot} G_\mathrm{TIA}
    \right)^2
    \int_0^B S_\mathrm{rel}(f)\,df
    } .
\end{equation}

If the relative noise is approximately white over the detection bandwidth, the integral can be written as
\begin{equation}
    \int_0^B S_\mathrm{rel}(f)\,df
    \approx
    S_\mathrm{rel} B ,
\end{equation}
where $S_\mathrm{rel}$ has units of $\mathrm{Hz}^{-1}$. In this case,
\begin{equation}
    \mathrm{SNR}
    =
    \frac{B}{\mathrm{RBW}} \frac{
    G_\mathrm{TIA}^2 \mathcal{R}^2
    \beta^2 P_\mathrm{LO} P_\mathrm{0}
    }{
    S_\mathrm{V, el}^2 B
    +
    2 e \mathcal{R} P_\mathrm{tot} G_\mathrm{TIA}^2 B
    +
    \left(
    \mathcal{R} P_\mathrm{tot} G_\mathrm{TIA}
    \right)^2
    S_\mathrm{rel} B
    } .
\end{equation}
We now define the modulation efficiency $\beta^2_p$ by writing the modulation index as
\begin{equation}
    \beta^2 = \beta_p^2 P_\mathrm{m},
\end{equation}
where $P_\mathrm{m}$ is the incident mmWave/THz power and $\beta_p^2$ is the modulation efficiency in units of $\mathrm{W}^{-1}$. In this form, the SNR becomes
\begin{equation}
    \mathrm{SNR}
    =
    \frac{B}{\mathrm{RBW}} \frac{
    G_\mathrm{TIA}^2 \mathcal{R}^2
    \beta_p^2 P_\mathrm{m} P_\mathrm{LO} P_\mathrm{0}
    }{
    S_\mathrm{V, el}^2 B
    +
    2 e \mathcal{R} P_\mathrm{tot} G_\mathrm{TIA}^2 B
    +
    \left(
    \mathcal{R} P_\mathrm{tot} G_\mathrm{TIA}
    \right)^2
    S_\mathrm{rel} B
    } .
\end{equation}

The noise-equivalent power, $\mathrm{NEP}$, is defined as the input microwave power $P_\mathrm{m}$ for which $\mathrm{SNR}=1$. Neglecting the power-independent electronic noise term $S_\mathrm{V, el}$, this gives
\begin{equation}
    \mathrm{NEP}
    =
    \frac{\mathrm{RBW}}{B}
    \frac{
    2 e \mathcal{R} P_\mathrm{tot} G_\mathrm{TIA}^2 B
    +
    \left(
    \mathcal{R} P_\mathrm{tot} G_\mathrm{TIA}
    \right)^2
    S_\mathrm{rel} B
    }{
    G_\mathrm{TIA}^2 \mathcal{R}^2
    \beta_p^2 P_\mathrm{LO} P_\mathrm{0}
    } .
\end{equation}
Since both noise terms scale linearly with the detection bandwidth $B$, the bandwidth cancels, yielding
\begin{equation}
    \mathrm{NEP}
    =
    \mathrm{RBW}
    \frac{
    2 e \mathcal{R} P_\mathrm{tot} G_\mathrm{TIA}^2
    +
    \left(
    \mathcal{R} P_\mathrm{tot} G_\mathrm{TIA}
    \right)^2
    S_\mathrm{rel}
    }{
    G_\mathrm{TIA}^2 \mathcal{R}^2
    \beta_p^2 P_\mathrm{LO} P_\mathrm{0}
    } .
\end{equation}
Equivalently, this can be written as
\begin{equation}
    \mathrm{NEP}
    =
    \frac{\mathrm{RBW}}{\beta_p^2 P_\mathrm{LO} P_\mathrm{0}}
    \left[
    \frac{2 e P_\mathrm{tot}}{\mathcal{R}}
    +
    P_\mathrm{tot}^2 S_\mathrm{rel}
    \right] .
\end{equation}

The first term corresponds to the shot-noise-limited contribution, while the second term corresponds to relative intensity-like noise. Therefore, the shot-noise-limited NEP is
\begin{equation}
    \mathrm{NEP}_\mathrm{shot}
    =
    \frac{\mathrm{RBW}}{\beta_p^2 P_\mathrm{LO} P_\mathrm{0}}
    \frac{2 e P_\mathrm{tot}}{\mathcal{R}},
\end{equation}
whereas the relative-noise-limited contribution is
\begin{equation}
    \mathrm{NEP}_\mathrm{rel}
    =
    \frac{\mathrm{RBW}}{\beta_p^2 P_\mathrm{LO} P_\mathrm{0}}
    P_\mathrm{tot}^2 S_\mathrm{rel}.
\end{equation}

The displayed average noise level (DANL) is obtained by normalizing the NEP to the resolution bandwidth:
\begin{equation}
    \mathrm{DANL}
    =
    \frac{\mathrm{NEP}}{\mathrm{RBW}}.
\end{equation}
Thus,
\begin{equation}
    \mathrm{DANL}
    =
    \frac{1}{\beta_p^2 P_\mathrm{LO} P_\mathrm{0}}
    \left[
    \frac{2 e P_\mathrm{tot}}{\mathcal{R}}
    +
    P_\mathrm{tot}^2 S_\mathrm{rel}
    \right] .
\end{equation}
To independently estimate the noise parameters used in the DANL model, we measured the photodetector output-noise variance as a function of the total detected optical power $P_\mathrm{tot}$. Without an EDFA, the measured variance follows a linear dependence on optical power, consistent with shot-noise-limited photodetection:
\begin{equation}
    V_\mathrm{noise}^2
    =
    S_\mathrm{V, el}^2 B
    +
    2 e \mathcal{R} P_\mathrm{tot} G_\mathrm{TIA}^2 B .
\end{equation}
From the slope of this linear fit, we extract the effective transimpedance gain $G_\mathrm{TIA}$. The fitted offset $V_0^2$ is small compared with the optical shot-noise contribution over the relevant power range, confirming that the power-independent electronic noise can be neglected in the DANL estimate.

When the EDFA is inserted before the photodetector, the noise variance exhibits a clear superlinear dependence on $P_\mathrm{tot}$. We fit this dataset using
\begin{equation}
    V_\mathrm{noise}^2
    =
    S_\mathrm{V, el}^2 B
    +
    2 e \mathcal{R} P_\mathrm{tot} G_\mathrm{TIA}^2 B
    +
    \left(
        \mathcal{R} P_\mathrm{tot} G_\mathrm{TIA}
    \right)^2
    S_\mathrm{rel} B ,
\end{equation}
where $G_\mathrm{TIA}$ is fixed to the value extracted from the shot-noise-limited measurement. The only additional power-dependent parameter is therefore the effective relative-noise spectral density $S_\mathrm{rel}$, which captures EDFA-induced intensity noise together with other relative intensity-like technical noise sources.

\begin{figure}[h!]
    \centering
     \includegraphics[width=0.5\linewidth]{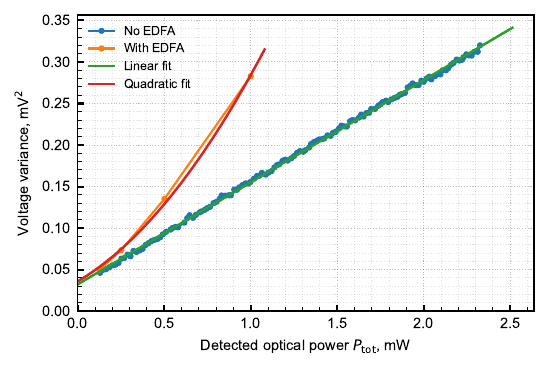}
    \caption{
    Photodetector noise calibration used for the DANL estimate. The voltage-noise variance is measured as a function of the total detected optical power $P_\mathrm{tot}$. Without EDFA, the noise follows a linear dependence on optical power, confirming shot-noise-limited detection. With EDFA, the variance follows a quadratic dependence due to excess relative intensity-like noise. The quadratic coefficient is used to extract the effective relative-noise spectral density $S_\mathrm{rel}$, which is then used in the DANL calculation.
    }
    \label{fig:pd_noise}
\end{figure}

The measured EDFA-assisted noise is well reproduced by the quadratic model, allowing us to retrieve $S_\mathrm{rel}$ directly from the experimental data:
\begin{equation}
    S_\mathrm{rel}
    =
    \frac{a}{\mathcal{R}^2 G_\mathrm{TIA}^2 B},
\end{equation}
where $a$ is the fitted quadratic coefficient expressed in SI units of $\mathrm{V^2/W^2}$. The extracted value of $S_\mathrm{rel}=-154.40\pm0.23$ dBc/Hz is then used in the DANL model without additional fitting parameters. Using simulated modulation efficiencies $\beta^2/P_\mathrm{m}$ in the range $8\cdot10^{-2}~\mathrm{W}^{-1}$, the calculated DANL is expected to be -105.8 dBm/Hz fitting well with the experimentally measured sensitivity presented in the Figure 2 of the main text. This agreement confirms that the observed detection limit is consistent with independently measured photodetector noise and EDFA-induced relative intensity noise. 

\section{Additional details on the phase and amplitude noise measurements}\label{SInote3}

\subsection{Signal-to-noise ratio for pulsed electro-optic sampling}\label{si3_snr_pulsed}

The incident mmWave or terahertz field induces a small electro-optic modulation of the optical pulse train. For a sinusoidal mmWave signal, we write the optical power incident on the photodiode as
\begin{equation}
    P_\mathrm{PD}(t)
    =
    P_\mathrm{0}
    \left[
        1-\beta\cos\left(\Omega t\right)
    \right].
\end{equation}
The photodiode current is
\begin{equation}
    i(t)
    =
    \mathcal{R} P_\mathrm{PD}(t),
\end{equation}
and the voltage at the output of the transimpedance amplifier is
\begin{equation}
    V(t)
    =
    G_\mathrm{TIA}\mathcal{R}P_\mathrm{PD}(t).
\end{equation}

Because the optical pulse train samples the modulation once every $T_\mathrm{rep}$, the mmWave/THz frequency $f$ is mapped to an aliased frequency
\begin{equation}
    f_\mathrm{IF}
    =
    \left| f-k f_\mathrm{rep} \right|,
\end{equation}
where $k$ is an integer harmonic of the pulse repetition rate. The useful voltage component at the aliased frequency has RMS amplitude
\begin{equation}
    V_\mathrm{sig}
    =
    \frac{
    G_\mathrm{TIA}\mathcal{R}P_\mathrm{0}\beta
    }{\sqrt{2}} .
    \label{eq:pulsed_vsig}
\end{equation}
Here the factor $\sqrt{2}$ converts the sinusoidal voltage amplitude to an RMS value.
The total voltage-noise variance at the photodetector output, integrated over a detection bandwidth $B$, is
\begin{equation}
    V_\mathrm{noise}^2
    =
    S_\mathrm{V,el} B
    +
    2 e \mathcal{R} P_\mathrm{0} G_\mathrm{TIA}^2 B
    +
    \left(
        \mathcal{R}P_\mathrm{0}G_\mathrm{TIA}
    \right)^2
    \int_0^B S_\mathrm{rel}(f)\,df ,
    \label{eq:pulsed_noise}
\end{equation}
which is the same as in~\ref{noise_model_si2}. The power signal-to-noise ratio is defined as
\begin{equation}
    \mathrm{SNR}
    =
    \frac{V_\mathrm{sig}^2}{V_\mathrm{noise}^2}.
\end{equation}
Using Eq.~\eqref{eq:pulsed_noise}, this gives
\begin{equation}
    \mathrm{SNR}
    =
    \frac{
    G_\mathrm{TIA}^2\mathcal{R}^2P_\mathrm{0}^2\beta^2/2
    }{
    S_\mathrm{V,el} B
    +
    2 e \mathcal{R} P_\mathrm{0} G_\mathrm{TIA}^2 B
    +
    \left(
        \mathcal{R}P_\mathrm{0}G_\mathrm{TIA}
    \right)^2
    \int_0^B S_\mathrm{rel}(f)\,df
    } .
    \label{eq:pulsed_snr_basic}
\end{equation}
If the relative noise is approximately white over the detection bandwidth, the integral can be written as
\begin{equation}
    \int_0^B S_\mathrm{rel}(f)\,df
    \approx
    S_\mathrm{rel} B ,
\end{equation}
And neglecting the power independent electronic noise term  $S_\mathrm{V, el}$, we get:
\begin{equation}
    \mathrm{SNR}
    =
    \frac{1}{2B}
    \frac{
    \mathcal{R}P_\mathrm{0}\beta_p^2 P_m
    }{
    2 e 
    +
        \mathcal{R}P_\mathrm{0} S_\mathrm{rel}
    } .
    \label{eq:pulsed_snr_final}
\end{equation}
where $P_\mathrm{m}$ is the incident mmWave/THz power and $\beta_p^2$ is the modulation efficiency in units of $\mathrm{W}^{-1}$. The prefactor in this formula is linked to the acquisition time $T_\mathrm{acq}$ through $T_\mathrm{acq} = \frac{1}{2B}$. Equation~\eqref{eq:pulsed_snr_final} separates the two optical-noise contributions in a compact form. The term $2e$ in the denominator corresponds to optical shot noise, whereas the term $\mathcal{R}P_0S_\mathrm{rel}$ corresponds to relative intensity-like noise. In the shot-noise-limited regime, where
\begin{equation}
    \mathcal{R}P_0S_\mathrm{rel} \ll 2e,
\end{equation}
the SNR scales linearly with the detected optical power,
\begin{equation}
    \mathrm{SNR}_\mathrm{shot}
    \approx
    \frac{1}{2B}
    \frac{
    \mathcal{R}P_0\beta_p^2P_m
    }{
    2e
    } .
\end{equation}
In contrast, when the relative intensity-like noise dominates,
\begin{equation}
    \mathcal{R}P_0S_\mathrm{rel} \gg 2e,
\end{equation}
the SNR becomes
\begin{equation}
    \mathrm{SNR}_\mathrm{rel}
    \approx
    \frac{1}{2B}
    \frac{
    \beta_p^2P_m
    }{
    S_\mathrm{rel}
    } .
\end{equation}
In this limit, increasing the optical power no longer improves the SNR, because both the signal and the relative intensity-like noise scale proportionally with $P_0$. In our measurements, the EDFA was placed after the chip to increase the optical power incident on the photodetector and to operate in the regime where the power-dependent optical noise dominates over the power-independent electronic noise. 

\subsection{I/Q demodulation referenced to the repetition rate of the optical frequency comb}

The recorded time-domain signal was processed using a self-developed Python routine. First, the dominant intermediate-frequency beat note $f_{\mathrm{IF}}$ was estimated from the Fourier spectrum of a short section of the recorded trace. 
This frequency was then used as the digital reference for I/Q demodulation of the full recording. 
For a measured real-valued signal $s(t)$, the in-phase and quadrature components were obtained by mixing the signal with cosine and sine reference waveforms at $f_{\mathrm{IF}}$,
\begin{equation}
    I_{\mathrm{raw}}(t) = 2 s(t)\cos\left(2\pi f_{\mathrm{IF}} t\right),
    \qquad
    Q_{\mathrm{raw}}(t) = -2 s(t)\sin\left(2\pi f_{\mathrm{IF}} t\right).
\end{equation}
The factor of two accounts for the conversion from a real-valued passband signal to its complex baseband representation. 
After mixing, both quadratures were low-pass filtered to remove the components near $2f_{\mathrm{IF}}$, leaving the slowly varying baseband components $I(t)$ and $Q(t)$.

The filtered in-phase and quadrature components were combined into a complex baseband signal,
\begin{equation}
    z(t) = I(t) + iQ(t).
\end{equation}
The phase fluctuations were obtained from the unwrapped argument of this signal, while amplitude fluctuations were extracted from the normalized signal power,
\begin{equation}
    \phi(t) = \mathrm{unwrap}\left[\arg z(t)\right],
    \qquad
    A(t) = \frac{|z(t)|^2 - \langle |z(t)|^2 \rangle}{\langle |z(t)|^2 \rangle}.
\end{equation}
A residual linear phase ramp was subtracted from $\phi(t)$ to remove the remaining carrier-frequency offset. The phase and amplitude fluctuation spectra were then calculated using Welch averaging, yielding $S_{\phi}(f)$ and $S_A(f)$, respectively. Long recordings were processed block by block while preserving filter states and phase continuity between consecutive blocks.

In measurements where the amplitude modulation is known to be negligible, the recovered amplitude fluctuations provide an additional consistency check for the phase reconstruction. In particular, a stable amplitude trace confirms that the observed fluctuations are dominated by phase noise rather than residual amplitude-to-phase mixing or fitting artifacts. This criterion was used to verify the robustness of the phase-noise extraction.

\subsection{Phase-noise budget of comb-referenced aliasing}
\label{PN_budget}

In the comb-referenced measurement, a mmWave/THz carrier at frequency $f_{\mathrm{THz}}$ is sampled by the fs-laser pulse train with repetition rate $f_{\mathrm{rep}}$. The carrier is mapped to an aliased intermediate frequency

\begin{equation}
    f_{\mathrm{IF}} = \left| f_{\mathrm{THz}} - k f_{\mathrm{rep}} \right|,
    \label{eq:alias_frequency}
\end{equation}

where $k$ is the alias order. The same relation applies to the instantaneous phase. Neglecting the absolute sign, the measured aliased phase can be written as

\begin{equation}
    \phi_{\mathrm{IF}}(t)
    =
    \phi_{\mathrm{THz}}(t)
    -
    k \phi_{\mathrm{rep}}(t)
    +
    \phi_{\mathrm{ro}}(t),
    \label{eq:alias_phase}
\end{equation}

where $\phi_{\mathrm{THz}}(t)$ is the phase fluctuation of the mmWave/THz carrier, $\phi_{\mathrm{rep}}(t)$ is the phase fluctuation of the comb repetition rate, and $\phi_{\mathrm{ro}}(t)$ represents additional readout noise, including photodetection, digitizer-clock, and electronic acquisition noise.

Assuming these noise terms are uncorrelated, the measured phase-fluctuation power spectral density in the aliased domain is

\begin{equation}
    S_{\phi,\mathrm{IF}}(f)
    =
    S_{\phi,\mathrm{THz}}(f)
    +
    k^2 S_{\phi,\mathrm{rep}}(f)
    +
    S_{\phi,\mathrm{ro}}(f).
    \label{eq:alias_phase_noise_psd}
\end{equation}

Here, $S_\phi(f)$ denotes the one-sided phase-fluctuation power spectral density in $\mathrm{rad^2/Hz}$. The corresponding single-sideband phase noise is

\begin{equation}
    L(f)
    =
    10\log_{10}\left[\frac{S_\phi(f)}{2}\right],
    \label{eq:ssb_phase_noise_definition}
\end{equation}

with $L(f)$ expressed in $\mathrm{dBc/Hz}$. Equivalently, Eq.~\eqref{eq:alias_phase_noise_psd} can be written in single-sideband phase-noise notation as

\begin{equation}
    L_{\mathrm{IF}}(f)
    =
    10\log_{10}
    \left[
    10^{L_{\mathrm{THz}}(f)/10}
    +
    10^{\left(L_{\mathrm{rep}}(f)+20\log_{10}k\right)/10}
    +
    10^{L_{\mathrm{ro}}(f)/10}
    \right].
    \label{eq:alias_phase_noise_L}
\end{equation}

Thus, the measured aliased phase noise contains three contributions: the intrinsic phase noise of the mmWave/THz carrier, the comb repetition-rate phase noise multiplied by the alias order, and the readout-chain phase noise. The repetition-rate contribution to the aliased phase noise is

\begin{equation}
    L_{\mathrm{rep}\rightarrow\mathrm{IF}}(f)
    =
    L_{\mathrm{rep}}(f) + 20\log_{10} k .
    \label{eq:rep_noise_scaling}
\end{equation}

This scaling is important because $k \simeq f_{\mathrm{THz}}/f_{\mathrm{rep}}$ can be large for mmWave/THz carriers. Therefore, the comb repetition rate and the digitizer/readout clock must be sufficiently low-noise for the measured aliased phase noise to be dominated by the carrier under test rather than by the sampling reference.

For the measurements presented in the main text, Eq.~\eqref{eq:alias_phase_noise_L} provides the phase-noise budget for interpreting the aliased spectra. The measured phase noise agrees with the expected reference curves obtained from the RF seed after applying the $20\log_{10}(N)$ frequency-multiplication scaling, where $N$ is the multiplier-chain order. The direct 10.025~GHz validation measurement presented in Section~\ref{ESA_validation} further confirms that the comb repetition-rate contribution, estimated using Eq.~\eqref{eq:rep_noise_scaling}, remains below the measured carrier phase noise over the investigated offset-frequency range.

\subsection{Validation of amplitude-noise retrieval by THz on-off modulation}\label{AM_modulation}

\begin{figure}[h!]
    \centering
    \includegraphics[width=\linewidth]{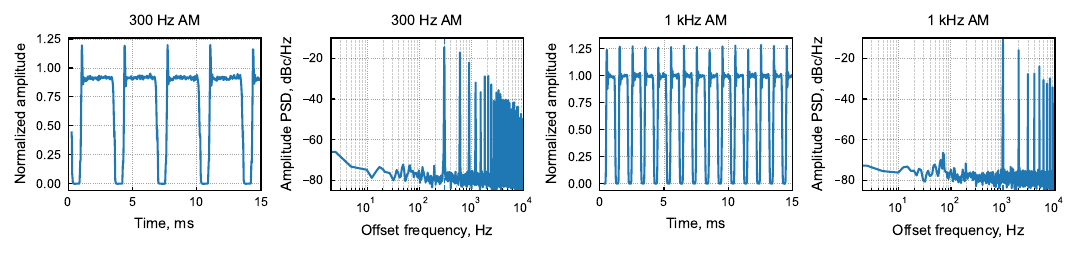}
    \caption{\textbf{Validation of amplitude-noise retrieval by applied THz amplitude modulation.}
    Time-domain amplitude traces and corresponding amplitude-noise spectra retrieved from the complex baseband signal after digital I/Q demodulation. Applied amplitude modulation at 300~Hz and 1~kHz produces clear peaks at set modulation frequencies, confirming that the same demodulation routine used for phase-noise extraction also recovers amplitude fluctuations of the mmWave/THz carrier.}
    \label{fig:AM_validation}
\end{figure}

To validate amplitude-noise retrieval, we applied square-wave amplitude modulation to the mmWave/THz carrier and processed the detected aliased signal using the same digital I/Q demodulation routine as for phase-noise extraction. The complex baseband signal $z(t)=I(t)+iQ(t)$ was used to calculate the normalized power fluctuation,
\begin{equation}
    A(t)=\frac{|z(t)|^2-\langle |z(t)|^2\rangle}{\langle |z(t)|^2\rangle}.
\end{equation}
As shown in Fig.~\ref{fig:AM_validation}, applied modulation at 300~Hz and 1~kHz is recovered in both the time-domain amplitude trace and the corresponding amplitude-noise spectrum. This confirms that the I/Q demodulation procedure retrieves amplitude fluctuations of the aliased carrier in addition to its phase fluctuations.
This measurement confirms that the I/Q demodulation procedure recovers not only the phase fluctuations of the aliased carrier, but also its amplitude dynamics. The corresponding amplitude-noise spectrum can therefore be obtained from $A(t)$ using Welch averaging, yielding the relative intensity noise spectrum discussed in the main text.

\subsection{Direct validation of phase-noise retrieval at 10.025~GHz}
\label{ESA_validation}

To validate the phase-noise retrieval independently of the mmWave frequency-multiplier chain, we performed a direct measurement at the RF seed frequency of 10.025~GHz. The same Keysight E8267D PSG Vector Signal Generator used in the main phase-noise measurements was operated at an output power of 23~dBm and radiated toward the TFLN receiver using a horn antenna with 10~dBi gain. On the receiver side, the realized antenna gain of the on-chip structure is expected to be approximately $-20$~dBi at this frequency. The resulting free-space coupling efficiency is therefore low, but still sufficient to recover the carrier and validate the phase-noise extraction procedure.

\begin{figure}[h!]
    \centering
    \includegraphics[width=\linewidth]{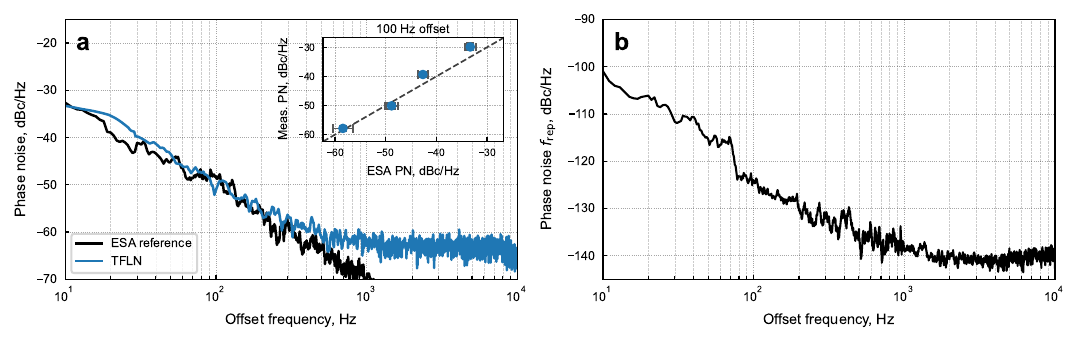}
    \caption{\textbf{Direct validation of phase-noise retrieval at 10.025~GHz.}
    \textbf{a,} Single-sideband phase noise measured with the TFLN receiver and with an electrical spectrum analyzer (ESA). Inset: TFLN-retrieved phase noise versus ESA reference at 100~Hz offset for different programmed phase-noise settings.
    \textbf{b,} Independently measured phase noise of the fs-laser repetition-rate signal. For the 10.025~GHz validation measurement, the alias order is $k\approx 100$, so the repetition-rate contribution to the aliased phase noise is obtained by adding $20\log_{10}k\approx 40$~dB. The scaled repetition-rate noise remains well below the measured RF-carrier phase noise, confirming that the validation measurement is not limited by the optical sampling clock.}
    \label{fig:10GHz_PN_validation}
\end{figure}

The detected aliased carrier was processed using the same digital I/Q demodulation routine as in the main text. Fig.~\ref{fig:10GHz_PN_validation}a compares the single-sideband phase-noise spectrum retrieved with the TFLN receiver to an independent measurement performed with an electrical spectrum analyzer (ESA). The agreement between the two spectra confirms that the TFLN receiver accurately recovers the phase noise of the 10.025~GHz carrier. The inset shows the TFLN-retrieved phase noise versus the ESA reference at a fixed offset frequency of 100~Hz for different programmed phase-noise settings, further confirming the linearity and quantitative accuracy of the phase-noise readout.

We also independently measured the phase noise of the fs-laser repetition-rate signal, shown in Fig.~\ref{fig:10GHz_PN_validation}b, to verify that the validation measurement is not limited by the optical sampling clock. According to Eq.~\eqref{eq:rep_noise_scaling}, the repetition-rate contribution to the aliased phase noise is obtained by adding $20\log_{10}k$ to the measured repetition-rate phase noise. For the present measurement, $f_{\mathrm{MW}}=10.025$~GHz and $f_{\mathrm{rep}}\approx 100$~MHz, giving $k\approx 100$ and therefore $20\log_{10}k\approx 40$~dB. At 100~Hz offset, the measured repetition-rate phase noise is approximately $L_{\mathrm{rep}}\approx -126$~dBc/Hz, corresponding to a scaled contribution of $L_{\mathrm{rep}\rightarrow\mathrm{IF}}\approx -86$~dBc/Hz. This is more than 35~dB below the measured 10.025~GHz carrier phase noise. These measurements show that the recovered phase noise is dominated by the MW carrier under test rather than by the comb repetition-rate noise.

\section{Additional details on the mutual coherence measurements}\label{si_note4}

Mutual-coherence measurements were performed using the same mmWave source as in the previous experiments, operated slightly outside its nominal single-tone operating band. Under these conditions, the source emitted several discrete spectral components, corresponding to different harmonic products of the RF drive. We first analyzed the emitted radiation using the heterodyne measurement technique described above in order to confirm its frequency content. Then, these components were detected simultaneously by the photonic chip pumped by the fs laser, and so the mmWave lines appeared as multiple aliased intermediate-frequency lines in the recorded time-domain signal.

For each observed line, the corresponding intermediate frequency $f_{\mathrm{IF},i}$ was first estimated from the Fourier spectrum of the recorded trace. The measured real-valued signal $s(t)$ was then digitally demodulated around each $f_{\mathrm{IF},i}$ using the same I/Q procedure described above,
\begin{equation}
    I_{\mathrm{raw},i}(t) = 2 s(t)\cos\left(2\pi f_{\mathrm{IF},i} t\right),
\end{equation}
\begin{equation}
    Q_{\mathrm{raw},i}(t) = -2 s(t)\sin\left(2\pi f_{\mathrm{IF},i} t\right).
\end{equation}
After low-pass filtering, the slowly varying quadratures $I_i(t)$ and $Q_i(t)$ were obtained and combined into the complex baseband signal
\begin{equation}
    z_i(t) = I_i(t) + iQ_i(t).
\end{equation}
The phase trajectory of each line was then extracted from the unwrapped argument of the corresponding complex baseband signal,
\begin{equation}
    \phi_i(t) = \mathrm{unwrap}\left[\arg z_i(t)\right].
\end{equation}
A residual linear phase ramp was subtracted from each $\phi_i(t)$ to remove the remaining carrier-frequency offset, yielding the phase fluctuation
\begin{equation}
    \delta\phi_i(t) = \phi_i(t) - 2\pi \delta f_i t - \phi_{0,i},
\end{equation}
where $\delta f_i$ is the residual frequency offset after digital demodulation and $\phi_{0,i}$ is a constant phase offset.

\end{document}